\documentclass[aps,prx,twocolumn,showpacs,floatfix,nobibnotes,nofootinbib,superscriptaddress,amsmath,amssymb,longbibliography]{revtex4-1}
\usepackage[english]{babel}
\usepackage[colorlinks,urlcolor={blue}]{hyperref}
\usepackage{eurosym}
\usepackage[usenames]{color}
\usepackage{graphicx}
\usepackage{amsmath}
\usepackage{amsfonts}
\usepackage{amssymb}
\usepackage{mathrsfs}
\usepackage{bm}
\usepackage{verbatim}
\usepackage{ulem}
\usepackage{lipsum}
\usepackage{braket}

\usepackage[textsize=small]{todonotes}

\newcommand{\beq}{\begin{equation}}
\newcommand{\eeq}{\end{equation}}
\newcommand{\bea}{\begin{align}}
\newcommand{\eea}{\end{align}}

\begin{document}

\title{\textbf{Feasibility of perturbative generation of bound-states from resonances or virtual states}}
\author{C.-J. Yang}
\affiliation{ELI-NP, ``Horia Hulubei" National Institute for Physics and Nuclear Engineering, 30 Reactorului Street, RO-077125, Bucharest-Magurele, Romania}
\email{chieh.jen@eli-np.ro}
\affiliation{Nuclear Physics Institute of the Czech Academy of Sciences, 
              25069 \v{R}e\v{z}, Czech Republic}

\begin{abstract}
I investigate whether it is possible to generate bound-states from
resonances or virtual states through first-order perturbation theory. 
Using contact-type
potentials as those appeared in pionless effective field theory, I show that it is possible to obtain
negative-energy states by sandwiching a next-to-leading order (NLO) interaction with the leading-order (LO) wavefunctions, under 
the presence of LO resonances or virtual states. However, at least under the framework of time-independent Schr\"odinger
equation and Hermitian Hamiltonian, there is an inability to create bound-states with structure similar to those formed by the non-perturbative treatments. 

%
\end{abstract}

\keywords{nuclear theory}
\maketitle

\section{Introduction}
It is well-known that bound-states correspond to poles in the S-matrix~\cite%
{rubin}. Under the condition that an interaction contains no pole itself,
non-perturbative treatments are required to give rise a pole in the final amplitude. In
non-relativistic quantum mechanics, this is usually achieved by treating the
interaction as a potential in the Schr\"odinger or Lippmann-Schwinger equation
(LSE). 
Meanwhile, it could occur that a consistent
treatment of interactions under certain frameworks of effective field theory (EFT) involves a non-perturbative treatment of the leading-order (LO)
potential, while higher-order corrections are to be added perturbatively
under distorted-wave-Born-approximation (DWBA)~\cite%
{vanKolck:2020llt,vanKolck:2020plz}. For example, this occurs in the
standard formulation of pionless EFT~\cite{Hammer:2019poc} and the modified chiral EFT~\cite%
{ksw,ksw1,Birse:2007sx,Long:2007vp,Valdper,Valdperb,BY,BYb,BYc,Wu:2018lai,Peng:2021pvo,Peng:2020nyz,Habashi:2021pbe,Thim:2023fnl,Li:2023hwe,Thim:2024yks,Thim:2024jdv} approaches to low-energy nuclear physics.
Since a LO
interaction corresponds to the first approximation in describing a
system, it could contain considerable theoretical uncertainty.
Under a power counting which prescribes a perturbative arrangement of sub-leading interactions, one demands
that the uncertainty of the final observables can always be improved order-by-order perturbatively.   
 
However, one could encounter a situation in which a bound-state needed to be restored from
an unbound LO result. For example, $%
^{16}$O is shown to be unstable with respect to decaying into four
alpha-particles under pionless EFT at LO~\cite{pionless16,pionless16b} and chiral EFT without three-body
forces up to NLO~\cite{Yang:2020pgi}. Although for chiral EFT, a promotion of three-body forces to LO is found based on 
a combinatorial argument, which fixes the problem of $^{16}$O binding~\cite{yang_rev,Yang:2021vxa}, it is still of interest to investigate whether it is possible to generate bound-states perturbatively from
resonances or virtual states, as this might be demanded if one adopts another theoretical framework or studies other systems. 
In fact, a recent study suggests
that any few-body system obtained by zero-range momentum-independent two-body
interactions is unstable against decay into clusters, if its two-body
scattering length is much larger than any other scale involved~\cite%
{Schafer:2020ivj}. Under that scenario, one must recover the correct pole structure corresponding
to physical bound-states through perturbative corrections, at least for systems where the EFT is applicable. 
However, it is not obvious that this can be achieved easily, 
as it involves a transition between eigen-states belonging to two different Hilbert spaces. 

Since poles are infinities in the S-matrix, they cannot be
generated from nothing through finite steps of perturbative corrections.
Controlled resummations (e.g., Refs.~\cite{Bissegger:2008ff,PhysRevC.67.055202}) can usually be justified under EFT to overcome this problem, provided that the interactions being resumed are free of the Wigner bound effect~\cite{wigner,wigner2}. However, integrating such resummations (to be applied on certain momenta) along with the perturbative treatment (to be applied elsewhere) is usually possible only when the amplitude can be solved analytically. Moreover, it quickly becomes infeasible if more than three particles are considered effectively as the degree of freedom of the system.
Therefore, it is desirable if one could perturbatively convert the resonances or virtual states---which are poles in the S-matrix with energies analytic continued to the complex plane---into bound-states directly.
Note that resonance or virtual state poles consist of complex energies and \textit{are not} eigen-values of a Hermitian Hamiltonian. 
Under the non-perturbative setting, one could analytic continue the Schr\"odinger equation or apply methods such as complex scaling~\cite{Aguilar1971,Balslev1971} or other alternatives~\cite{Horek2017,Dietz:2021haj} to extend the domain of the model space. 
Despite the technical difficulty which might prevent one from solving the multi-particle systems in practice, the analytic continued wavefunctions are standard and well-defined. 
However, conceptual issues arise once one enters the perturbative regime. 
First, the S-matrix becomes non-unitary when perturbation theory is applied, which forces the analytic continuation toward a dangerous foundation.
The judgement of perturbativity then unavoidably involves comparisons between complex-to-complex or complex-to-real numbers. 
Thus, the smallness of a perturbative correction---which justifies the convergence of a perturbation series---can be
difficult to define in general. 

In this work, I study the problem within the simplest, Hermitian formalism. It is of interest to check whether bound-states can be generated from resonances or virtual states by straightforwardly applying the first-order perturbation theory, as this is probably the only practical way to be applied numerically in an A-body system with A$\gtrsim4$. On the other hand, following the idea originated from lattice-gauge methods~\cite{Symanzik:1983dc,Symanzik:1983gh,Lee:2008fa}, an indirect solution would be to avoid the problem by treating some of the subleading corrections non-perturbatively to ``improve" the LO convergence~\cite{Contessi:2023yoz,Contessi:2024vae}. 

I first layout the methodology of
calculation and the condition whether a first-order perturbative correction
can be considered as next-to-leading (NLO) under an EFT expansion in Sec. %
\ref{method}. Then, in Sec. \ref{3p0}, I investigate whether a resonance
can be converted to bound-states using the nucleon-nucleon (NN) $^{3}$P$_{0}$ channel as a testing ground. 
In Sec. \ref{1s0}, I investigate whether a virtual state can be
converted to bound-states using the NN $^{1}$S$_{0}$ channel as a testing ground.
Then, I discuss a reverse procedure regarding the removal of a LO bound-state and its implications in Sec. \ref{rev},
with a numerical example presented in Sec. \ref{rev_num}. 
The essence of the problem is discussed in Sec. \ref{avoid}.
Finally, I summarize the results and skepticize a generalization to
many-body systems in Sec. \ref{summary}.

\section{Methodology}

\label{method} 

\subsection{The DWBA formalism}
In this work all interactions are assumed to be finite-ranged and go to zero in the coordinate space at $r\rightarrow \infty$.
In fact, the problem will be analyzed in the momentum space with interactions that decay exponentially after $p\gg \Lambda$, where $\Lambda$ is a
momentum cutoff.  
I will always solve the time-independent Schr\"odinger or Lippmann-Schwinger equation to generate the LO results
non-perturbatively, and then add subleading corrections under DWBA. 
Denoting $p$ $(p^{\prime})$ the incoming (outgoing) momentum in the
center of mass (c.m.) frame of two equal-mass fermions with mass $M$ ($=939$ MeV in this work), the LO T-matrix regarding their scattering process 
in an uncoupled partial-wave can be obtained by
iterating the LO potential $V_{LO}(p,p^{\prime})$ non-perturbatively in the
LSE, i.e., 
\begin{widetext}
\begin{equation}
T_{LO}(p^{\prime },p;E)=V_{LO}(p^{\prime
},p)+\frac{2}{\pi }M\int_{0}^{\infty }\frac{
dp^{\prime \prime }\;p^{\prime \prime }{}^{2}\;V_{LO}(p^{\prime },p^{\prime \prime })\;T_{LO}(p^{\prime \prime
},p,E)}{p_{0}^{2}+i\varepsilon -p^{\prime \prime }{}^{2}},
\label{eq:2.3}
\end{equation}
\end{widetext}
where $p_{0}^{2}/M=E_{c.m.}$ is the c.m. energy. Note that the spin, isospin and angular momentum labels are dropped in Eq.~(\ref{eq:2.3}), as only uncoupled partial-waves will be evaluated in this work. 
To simplify the notation, I rewrite Eq.~(\ref{eq:2.3}) into the following operator form~\footnote{$T$, $G$ and $V$ are to be understood as operators, just like those appear in e.g., Refs.~\cite{Yang:2007hb,Yang:2009kx,Yang:2009kx}. }
\begin{equation}
T_{LO}=V_{LO}+V_{LO}GT_{LO},\label{aa}
\end{equation}
with
\begin{equation}
G=\frac{2}{\pi }\frac{M}{p_{0}^{2}+i\varepsilon -p^{\prime \prime }{}^{2}}.
\end{equation}
Note that the integrals that are implicit in Eq.~(\ref{aa}) (and the following) are
one-dimensional integrals, since results are being computed partial-wave by partial-wave. The NLO contribution is then given by DWBA and reads~\cite{BY} 
\begin{align}
T_{NLO}& =V_{NLO}+V_{NLO}GT_{LO}+T_{LO}GV_{NLO} \nonumber\\
& + T_{LO}G\,V_{NLO}\,GT_{LO}\, .   \label{eqn:LSE23}
\end{align}
where $V_{NLO}$ and $T_{NLO}$ are the interaction and T-matrix at NLO. The
LO phase shift $\delta_{LO}$ is related to the LO on-shell T-matrix by 
\begin{equation}
\frac{e^{i\delta_{LO}}\sin\delta_{LO}}{-M p_0}=T_{LO}(p_0,p_0;E).
\label{lophase}
\end{equation}
Meanwhile, phase shifts up to NLO are in principle to be obtained via a
perturbative conversion, since the S-matrix corresponds to $T_{LO}+T_{NLO}$
does not in general present the unitary property. Under the condition that
unitarity can be restored order-by-order, the NLO phase shifts $\delta_{NLO}$
can be obtained from $\delta_{LO}$ and $T_{NLO}$ by~\cite{BY} 
\begin{equation}
T_{NLO}(p_0,p_0;E_{c.m.})=-\delta_{NLO}\frac{e^{2i\delta_{LO}}}{-M p_0}.
\label{nlophase}
\end{equation}
On the other hand, Eq.~(\ref{nlophase}) cannot be trusted if it results a $%
\delta_{NLO}$ far from a real number, which then put the perturbative
setting of power counting doubtable in the first place. Once the on-shell
T-matrix (at LO or NLO) is obtained, cross section $\sigma$ can be
calculated, i.e.,
\begin{equation}
\sigma=4\pi(2l+1)M^2|T|^2,  \label{cross}
\end{equation}
where $l$ is the angular momentum quantum number of the system and $T=T_{LO}$ ($T_{LO}+T_{NLO}$)
for order up to LO (NLO).

%

%

The leading perturbative correction in energy due to $V_{NLO}$ can be easily evaluated by sandwiching it with
the LO wavefunctions.
In this work the Harmonic-Oscillator (HO) basis is adopted to
diagonalize the LO Hamiltonian 
\begin{equation}
H_{LO}=T+V_{LO}, \label{lo}
\end{equation}%
where $T$ denotes the kinetics term. Denoting the $i^{th}$ LO eigen-value and
eigen-function as $E_{i}^{LO}$ and $\Psi _{i}^{LO}$, the NLO correction $%
E_{i}^{NLO}$ is then 
\begin{equation}
E_{i}^{NLO}=\langle \Psi _{i}^{LO}|V_{NLO}|\Psi _{i}^{LO}\rangle ,
\label{enlo}
\end{equation}%
where 
\begin{equation}
\Psi _{i}^{LO}=\sum_{n=0}^{n_{max}}C_{in}\psi _{n}^{HO}, \label{10}
\end{equation}%
and $C_{in}$ denotes the linear combination coefficient of the HO wavefunction $\psi _{n}^{HO}$ up to
principle quantum number $n=n_{max}$. One can then check if there is any
state with negative energy being generated at NLO, i.e., whether $E_{i}^{LO}+E_{i}^{NLO}<0$.

Note
that there are two caveats in the above approach. 
First, strictly speaking, Eq.(\ref{enlo}) is the exact first order perturbative correction of $V_{NLO}$. While in EFT, NLO corrections are associated with an uncertainty up to $O$(NNLO) and \textit{do not} have an exact value\footnote{This issue is crucial for the implementation of RG-invariant power countings in chiral EFT, and will be discussed along with some recent critics~\cite{Gasparyan:2022isg} in a future work~\cite{res_epel}.}. Thus, even a test based on the above procedure shows that a straightforward perturbative threshold-crossing is problematic, it does not rule out the possibility that adding small additional terms of order $O$(NNLO)---which is allowed by EFT---could overcome the problem.  
Second, the LO
eigen-functions---which are scattering waves---are approximated in terms of HO basis. These two
wavefunctions have different asymptotic behavior and a transition from one
to another only exists in the limit $n_{max}\rightarrow \infty $ and under
an oscillator strength of the HO function $\hbar \omega \rightarrow 0$.
Thus, any unbound to bound transition needed to be analyzed carefully by
reducing $\hbar \omega $ and increasing $n_{max}$. 
Nevertheless, as long as $V_{NLO}$ is finite-range, one only needs to resolute the LO scattering waves
by the HO basis within the range of $V_{NLO}$ to obtain reliable results of Eq.(\ref{enlo}).   
This can be checked by
increasing $n_{max}$ and decreasing $\hbar\omega$ until a convergence is reached. In this work, all results under a given $\hbar\omega$ are obtained by increasing $n_{nmax}$ until all of the eigen-energies converge within $1\%$. Relation between $\hbar\omega$, $n_{max}$ adopted and the corresponding ``ultraviolet" and ``infrared" cutoff can be found in Refs.~\cite{Stetcu:2006ey,PhysRevA.76.063613,Stetcu:2009ic,PhysRevC.87.044326,PhysRevC.89.044301,PhysRevC.90.064007,PhysRevC.86.054002}.

\subsection{An essential difference between perturbative and non-perturbative approaches}
When interactions are encoded into potentials, one common expectation is that if $V_{NLO}$ is small compared to $V_{LO}$, the corrections in the final amplitude should be small also, regardless of how it is obtained. 
Meanwhile, there is an essential difference between
the perturbative and non-perturbative approaches.

Within perturbation theory, an eigen-value always consists of its previous order value plus the new corrections,
and one can keep track of the evolution of each eigen-state order-by-order. This gives rise to
the possibility of tracking the phenomenon of level-crossing---i.e., a lower state can be shifted to a higher state 
so that a reordering of the energy spectrum occurs. Moreover, the gap between two eigen-states naturally goes through zero when two levels cross each other.

On the other hand, the non-perturbative procedure involves re-diagonalization whenever the potential is changed, and one can only
infer how each state evolves.
Moreover, it often comes with the so-called
avoided-level-crossing feature. As the interacting strength is increased to a critical value, 
two eigen-states which were initially drawn closer will experience a ``sudden jump" and exchange their positions 
with the gap between them remains finite\footnote{See for example, Ref.\cite{Von_Neumann1993-tp} or p.305 of Ref.\cite{Landau1981-lx}
for more detail.}. 

Note that the usual expectation regarding what is ``smallness" fails when one approaches the critical region of avoided-level-crossing.
To illustrate this point, one can consider a two-level system. At LO, two eigen-states are produced non-perturbatively by an interaction where the strength is just below the critical value. 
Then, any small correction on top of the LO interaction will produce a profound difference depending on how it is included. 
On the one hand, as long as the increase of the interaction strength is small, the correction should be linear, and first-order perturbation theory should apply. This will produce a correction that is indeed small and further reduce the gap between
the two states. On the other hand, a non-perturbative treatment will trigger the avoided-crossing feature
and increase the gap. Thus, normal intuitions no longer hold, and one faces the choice between
the perturbative and non-perturbative treatments. Ultimately, the choice relies on whether one trusts the expansion in terms of
the potentials or the final amplitude. The former calls for a non-perturbative treatment of the interactions, and the latter demands treating
corrections perturbatively. Under the condition that one understands the ultimate interaction or even can produce and fine-tune such potentials experimentally\footnote{One example will be the cold atom experiments~\cite{Madison2012}.}, the non-perturbative treatment should be adopted. Conversely, if the interaction is unclear, then in the spirit of EFT---where the expansion should be
arranged order-by-order in terms of the final amplitude---a perturbative treatment is a more natural choice. 

It needs to be stressed that any further iteration of a perturbative correction \textit{needs not} to converge order-by-order toward its non-perturbative correspondence for the perturbative approach to make sense. In fact, a non-perturbative treatment corresponds to adopting a particular set of irreducible diagrams and iterating them to all order, while ignoring all other quantum correction diagrams completely. It is possible that this treatment is inconsistent in the first place.
For example, without the entrance of new higher-order diagrams or counter terms, further iterations of $V_{NLO}$ in Eq.(\ref{enlo}) could generate non-renormalizable results~\cite{Long:2007vp,vanKolck:2020llt,Griesshammer:2021zzz,vanKolck:2021rqu}, which has been observed also under the presence of finite-range potentials in the context of chiral EFT potentials~\cite{Nogga:2005hy,PhysRevC.74.054001,PhysRevC.74.064004,Yang:2009kx,Yang:2009kx,Yang:2009pn,Zeoli2013-ro}. 

However, this also implies
that there is an essential difference regarding how eigen-energies are re-distributed between the two approaches, and 
therefore posts a potential difference between bound-states generated non-perturbatively and bound-states shifted across threshold by perturbation theory. This problem will be encountered in the later sections.

\section{Resonances to bound-states}

\label{3p0} In this section, I investigate the possibility of perturbatively
converting a resonance to bound states, using the nucleon-nucleon (NN) $^{3}$P$_{0}$ scattering as a testing ground.
A resonance appears when particles are
bound temporary and decayed at a later time, which corresponds to a pole in
the scattering amplitude with the energy or momentum being analytic
continued to the complex plane. It often occurs when an attractive potential
has a repulsive ``lip" or barrier above $E=0$---due to the angular momentum
barrier $l(l+1)/Mr^{2}$---that helps confine the particle~\cite{rubin,Li:2023hwe}. 

Breit-Wigner formula can be used to describe the amplitude $f$ near a resonance,
i.e., 
\begin{equation}
f=e^{i\delta }\sin \delta =\frac{\Gamma /2}{E_{R}-E-i\Gamma /2},  \label{bw}
\end{equation}%
where $E_{R}$ and $\Gamma $ are the energy and width of the resonance.

In
this work, I utilize the so-called Argand plot to confirm and extract $E_{R}$ and $\Gamma $ of a resonance.

To generate resonances in the NN $^3$P$_0$ channel,  Iadopt 
\begin{align}
V_{LO}(p,p')=\left[c_{lo}pp^{\prime}+d_{lo}pp^{\prime}(p^{2}+p^{\prime}{}^2)%
\right]f_R,  \nonumber \\
V_{NLO}(p,p')=\left[c_{nlo}pp^{\prime}+d_{nlo}pp^{\prime}(p^{2}+p^{\prime}{}^2)%
\right]f_R,  \label{r1}
\end{align}
where $c_{lo}$, $c_{nlo}$, $d_{lo}$ and $d_{nlo}$ are low-energy constants (LECs), and $%
V_{LO}$ ($V_{NLO}$) are to be solved non-perturbatively (perturbatively)
with a regulator 
\begin{equation}
f_R=exp\left[-\frac{p^4+p^{\prime}{}^4}{\Lambda^4}\right].  \label{r}
\end{equation}

\subsection{Case I}
\label{case1}
\begin{figure}[tbp]
\includegraphics[scale=0.3]{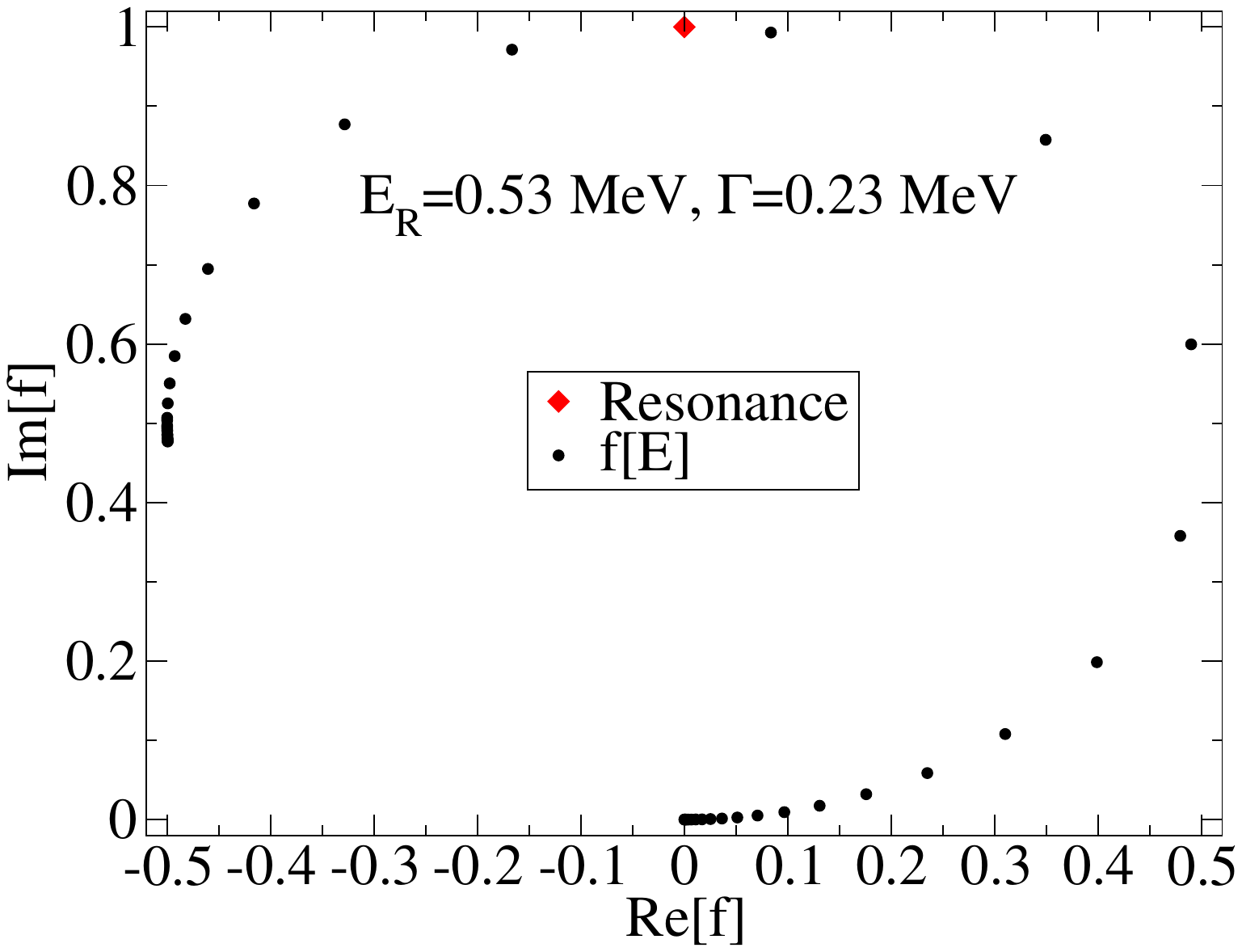}
\caption{Argand plot (Im[$f$] versus Re[$f$]) of the scattering amplitude $%
f(p_0)$ of case I. Starting at origin, an 1 MeV increase in $p_0=\protect%
\sqrt{ME}$ is applied to each point.
}
\label{ag3p0_c}
\end{figure}

\begin{figure}[tbp]
\includegraphics[scale=0.3]{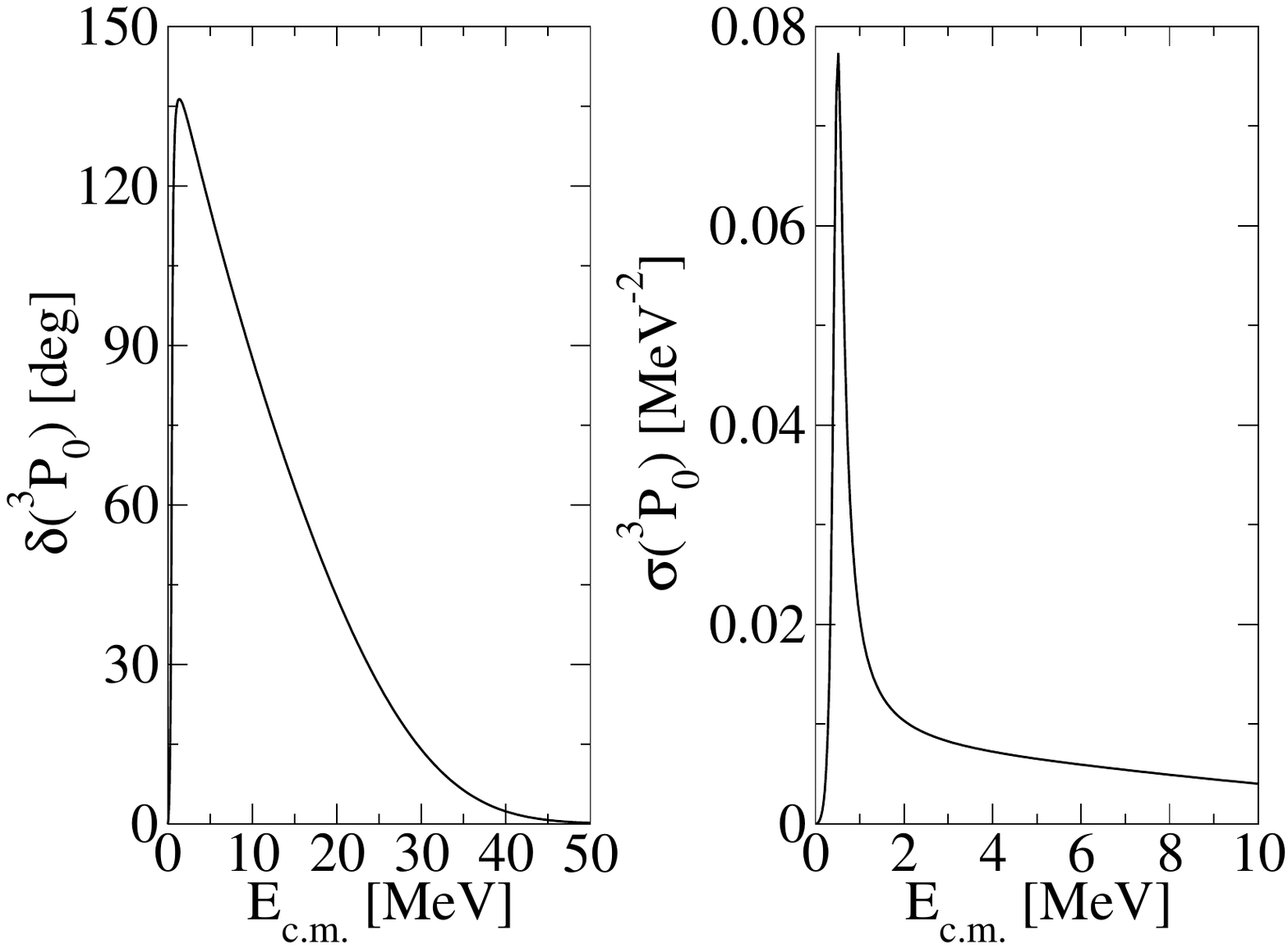}
\caption{Phase shifts $\protect\delta$ and cross section $\protect\sigma$ as
  a function of the c.m. energy $E_{c.m.}$ at LO of case I.  
}
\label{3p0_phase}
\end{figure}

 I first set $d_{lo}=0$ and $\Lambda=150$ MeV in Eq.~(\ref{r1}) and (\ref{r}). Inserting $c_{lo}=-2.5\times 10^{-9}$ MeV$^{-4}$ into Eq.~(%
\ref{eq:2.3}) (denoting as case I) gives an Argand plot illustrated in Fig.~%
\ref{ag3p0_c}. As the increase of $E=p_0^2/M$, the complex amplitude $f$
rotates counterclockwise around the circle. As the energy passes through the
resonance, $f$ becomes pure imaginary, and I extracted $E_R=0.53$ MeV and $\Gamma=0.23$ MeV according to
Eq.~(\ref{bw}). The corresponding phase shifts and differential cross
section are plotted as a function of $E_{c.m.}=T_{lab}/2$ in Fig.~\ref%
{3p0_phase}. Note that, instead of solving the LSE, one could also adopt the HO-basis to
diagonalize Eq.~(\ref{lo}), and then following the J-matrix method~\cite{Shirokov:2003kk,Shirokov:2004ff,Shirokov:2005bk,Yang:2016brl} to convert the eigen-energies
into scattering phase shifts. I have verified that the same $\delta(^3P_0)$ as listed in the left panel of Fig.~\ref{3p0_phase} can be
reproduced by using HO-basis with $\hbar \omega =1$ MeV
and $n_{max}=150$.

Next, the NLO correction is added. Note that solving LSE with a $10\%$ more attractive $c_{lo}$ (ie, $c_{lo}$ more
negative than $-2.8\times 10^{-9}$ MeV$^{-4}$) gives rise to an LO bound state. Thus,
to see how much extra effort is required to create a bound state
perturbatively against the non-perturbative way, only $c_{nlo}pp^{\prime }$ is adopted, which means $d_{nlo}=0$ in
Eq.~(\ref{r1}). Since the NLO corrections
are evaluated with Eq.~(\ref{enlo}), I first analysis how the results
behave with $\hbar \omega $ and $n_{max}$. For each given $\hbar \omega $,
the calculation is carried out by increasing $n_{max}$ until the resulting
energy spectra converge within $1\%$. Using $c_{lo}=-2.5\times 10^{-9}$ MeV$%
^{-4}$ and $c_{nlo}=-6.0\times 10^{-10}$ MeV$^{-4}$, $E^{LO}$ and $E^{NLO}$
obtained from different $\hbar \omega $ are compared in Table~\ref{t0}. As one
can see, perturbatively generated negative-energy states are observed when the wavefunctions are constructed by $\hbar
\omega \geq 4$ MeV, and disappear when $\hbar \omega =1$ MeV is
adopted. Those negative-energy states obtained with $\hbar\omega\geq 4$ MeV should therefore be discarded, as they are
merely an artifact of using bases consisting of bound-state wavefunctions with too large infrared truncation to describe scattering
waves. Nevertheless, one observes in Table~\ref{t0} that the
perturbative correction $E_{i}^{NLO}$ always peaked around the index $i$
where $E_{i}^{LO}\approx E_{R}=0.53$ MeV, i.e., the place where LO wavefunction inserted in Eq.~(%
\ref{enlo}) is closest to the resonance. Up to this point, two important
conclusions can be drawn. First, although based on the observed trend of
reducing $\hbar \omega $, one might always doubt whether an apparent NLO
bound state will disappear if $\hbar \omega $ is further reduced,
level crossing does occur. This is due to the fact that NLO corrections
peaked at $E_{R}$, which, when the continuum is discretized by
HO-wavefunction with small enough $\hbar \omega $, will always cause
states closer to the resonance to receive a larger correction than their
nearby states. Then, as long as this correction---which is negative and due to a tunable $%
V_{NLO}$---exceeds the increasing trend of $E_{i}^{LO}$, a state which was higher at LO will be
brought down more than its lower neighbors. Therefore, re-ordering of the
energy spectrum from LO to NLO will happen. One can further infer that
states with truly negative energies at NLO can be generated under a
combination of sufficiently attractive $V_{NLO}$ and closed-enough-to-zero $%
E_{R}$. However, since a resonance comes with a width $\Gamma $, in the
limit $\hbar \omega \rightarrow 0$, level crossing will not be limited
to only one state. Thus, at least under the setup of case I, continuous
states around $E_{R}$ will be brought to bound-states from any resonance
with a finite width. 

\begin{figure}[tbp]
\includegraphics[scale=0.3]{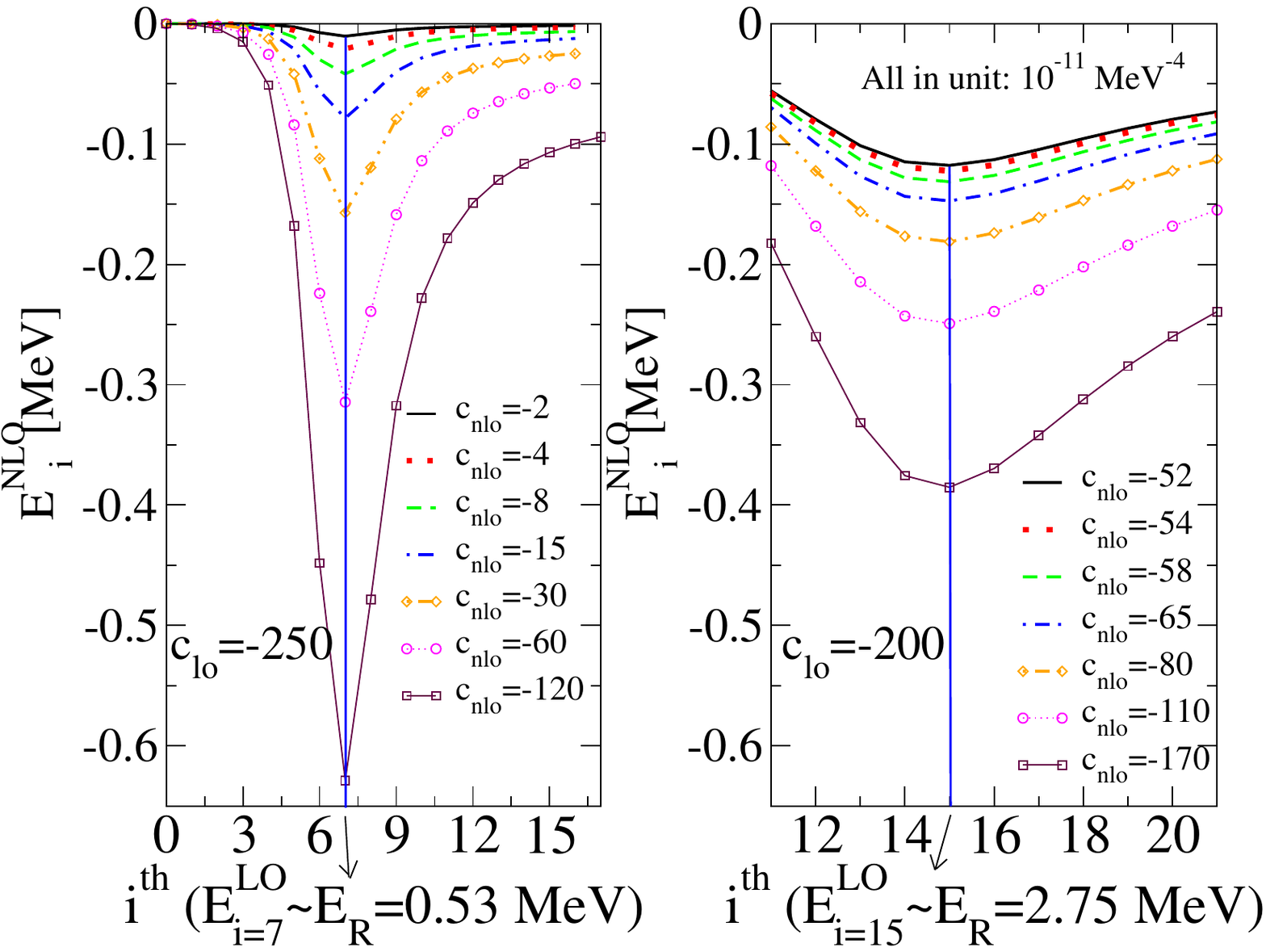}
\caption{$E^{NLO}_i$ obtained perturbatively with various $c_{nlo}$ under
$c_{lo}=-2.5\times 10^{-9}$ (left panel) and $c_{lo}=-2\times 10^{-9}$ (right panel) MeV$^{-4}$.
Note that $\Lambda=150$ MeV and $\hbar\protect\omega=1$ MeV are adopted; therefore, all the energies are discretized. The LO eigen-state which receives the largest NLO-correction is highlighted by the blue vertical line, and always corresponds to the index where $E_i^{LO}$ is the closest to $E_R$.}
\label{plot1_enlo}
\end{figure}

With these caveats in mind, one can examine how the NLO
results change with $c_{nlo}$. With $\hbar \omega =1$ MeV
and $n_{max}=150$, the NLO corrections is illustrated in the left panel of Fig.~\ref{plot1_enlo}.
 As expected,
the perturbative NLO correction $E^{NLO}_i$ is linear to $c_{nlo}$. One can then compare the results to
those obtained by treating NLO non-perturbatively.
Table~\ref{t1} shows that both the perturbative and non-perturbative generated
negative-energy states occur at NLO as the increase of $c_{nlo}$. In this particular
case, there is a 4 times difference between the $c_{nlo}$ required to
generate a NLO bound-state perturbative versus the case when it is treated
non-perturbatively. 
The corresponding cross sections $\sigma (^{3}P_{0})$ are plotted in Fig.~\ref%
{plot_cro}.

To see how the pole position of a resonance affects the NLO results, I adopt a less attractive $V_{LO}$ with $c_{lo}=-2\times 10^{-9}$ MeV$^{-4}$. 
This generates a broader resonance which is further away from the threshold with $E_{R}=2.75$ MeV and $\Gamma=2.52$ MeV.
The NLO correction are plotted in the right panel of Fig.~\ref{plot1_enlo}. Note that here I have adopted a more attractive $c_{nlo}$ so that
for the same legend in the left and right panels, the total strength $c_{lo}+c_{nlo}$ is the same.
As one can see, the perturbative corrections become weaker even with more attractive $c_{nlo}$. In fact, I am not able to obtain
any bound-state perturbatively up to NLO for all $c_{nlo}$ listed in the right panel of Fig.~\ref{plot1_enlo}.

\begin{table}[h]
\begin{tabular}{|c|c|c|c|c|c|c|}
\hline
i & \multicolumn{3}{|c|}{$\hbar \omega =20 $ MeV} & \multicolumn{3}{|c|}{$%
\hbar \omega =10$ MeV} \\ \hline
& $E_{i}^{LO}$ & $E_{i}^{NLO}$ & sum & $E_{i}^{LO}$ & $E_{i}^{NLO}$ & sum \\ 
\hline
0 & 0.48 & -0.98 & \textbf{-0.50} & 0.37 & -0.46 & \textbf{-9.20}$\times $%
\textbf{10}$^{-2}$ \\ \hline
1 & 1.34 & -0.58 & 0.76 & 0.73 & -0.79 & \textbf{-5.87}$\times $\textbf{10}$%
^{-2}$ \\ \hline
2 & 3.63 & -0.34 & 3.29 & 1.74 & -0.33 & 1.41 \\ \hline
3 & 7.20 & -0.28 & 6.92 & 3.37 & -0.24 & 3.13 \\ \hline\hline
i & \multicolumn{3}{|c|}{$\hbar \omega =4$ MeV} & \multicolumn{3}{|c|}{$%
\hbar \omega =1$ MeV} \\ \hline
& $E_{i}^{LO}$ & $E_{i}^{NLO}$ & sum & $E_{i}^{LO}$ & $E_{i}^{NLO}$ & sum \\ 
\hline
0 & 0.18 & -3.74$\times $10$^{-2}$ & 0.15 & 1.66$\times $10$^{-2}$ & -4.32$%
\times $10$^{-5}$ & 1.65$\times $10$^{-2}$ \\ \hline
1 & 0.44 & -0.58 & \textbf{-0.14} & 4.89$\times $10$^{-2}$ & -4.16$\times $10%
$^{-4}$ & 4.85$\times $10$^{-2}$ \\ \hline
2 & 0.73 & -0.54 & 0.19 & 9.71$\times $10$^{-2}$ & -2.01$\times $10$^{-3}$ & 
9.51$\times $10$^{-2}$ \\ \hline
3 & 1.31 & -0.26 & 1.05 & 0.16 & -7.52$\times $10$^{-3}$ & 0.15 \\ \hline
4 & 2.12 & -0.18 & 1.94 & 0.24 & -2.56$\times $10$^{-2}$ & 0.21 \\ \hline
5 & 3.16 & -0.15 & 3.01 & 0.33 & -8.41$\times $10$^{-2}$ & 0.24 \\ \hline
6 & 4.41 & -0.14 & 4.28 & 0.42 & -0.22 & 0.20 \\ \hline
7 & 5.88 & -0.12 & 5.76 & 0.52 & -0.31 & 0.21 \\ \hline
8 & 7.58 & -0.11 & 7.46 & 0.64 & -0.24 & 0.40 \\ \hline
\end{tabular}%
\caption{$E^{LO}_i$ and $E^{NLO}_i$ obtained with $\hbar\protect\omega%
=1,4,10,20$ MeV. Perturbatively generated bound-states are marked with bold
text. Results are obtained under case I with $c_{lo}=-2.5\times 10^{-9}$ MeV$%
^{-4}$ and $c_{nlo}=-6.0\times 10^{-10}$ MeV$^{-4}$. 
}
\label{t0}
\end{table}
\begin{table}[h]
\begin{tabular}{ccc}
\hline\hline
$c_{nlo} $ (MeV$^{-4}$) &  B.S.$_{per}$ (MeV) & B.S.$%
_{non-per}$ (MeV) \\ \hline
$-2\times 10^{-11}$  & --- & --- \\ 
$-4\times 10^{-11}$  & --- & --- \\ 
$-8\times 10^{-11}$ & --- & --- \\ 
$-15\times 10^{-11}$  & --- & --- \\ 
$-30\times 10^{-11}$ & --- & -0.194 \\ 
$-60\times 10^{-11}$ & --- & -1.09 \\ 
$-120\times 10^{-11}$  & -0.11, -0.026 & -3.19 \\ \hline\hline
\end{tabular}%
\caption{NLO results in $^3$P$_0$ channel under case I, obtained via
HO-wavefunctions with $\hbar\protect\omega=1$ MeV and $n_{max}=150$. 
B.S.$_{per}$ (B.S.$_{non-per}$) is the bound-state energy
obtained at NLO when treating Eq.~(\protect\ref{r1}) perturbatively
(non-perturbatively). ``---" denotes that no bound-state is generated. }
\label{t1}
\end{table}

\begin{figure}[tbp]
\includegraphics[scale=0.3]{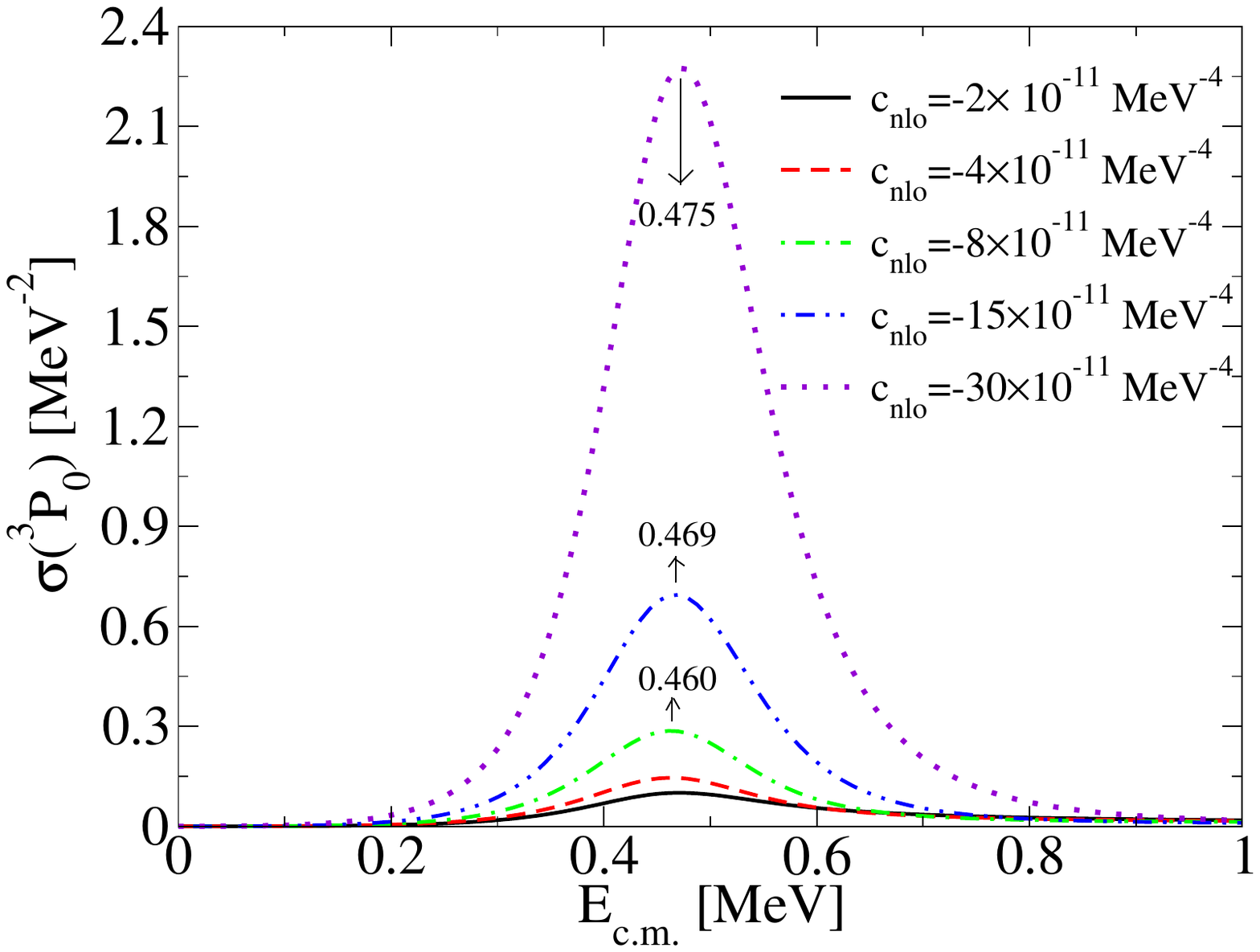}
\caption{Cross section $\protect\sigma$ as a function of c.m. energy $E_{c.m.}$
for treating $c_{nlo}=-2\sim-30\times 10^{-11}$ MeV$^{-4}$ perturbatively in Table~\protect\ref{t1}. The energy (in MeV) where $\protect\sigma$ peaked is explicitly given for the three largest $c_{nlo}$.}
\label{plot_cro}
\end{figure}

\subsection{Case II}
\label{case2}
Next, I include both the $c_{nlo}$ and $d_{nlo}$ terms in the NLO
interaction. Since their first-order perturbative corrections are linear to 
the prefactor $c_{nlo}$ or $d_{nlo}$, one could define
\begin{align}
C_{i}& =\langle \Psi _{i}^{LO}|pp^{\prime }|\Psi _{i}^{LO}\rangle , \\
D_{i}& =\langle \Psi _{i}^{LO}|pp^{\prime }(p^{2}+p^{\prime }{}^{2})|\Psi
_{i}^{LO}\rangle ,  \label{cd}
\end{align}%
and Eq.~(\ref{enlo}) can be rewritten as 
\begin{equation}
E_{i}^{NLO}=c_{nlo}C_{i}+d_{nlo}D_{i}.  \label{enlo2}
\end{equation}%
I now probe the possibility of creating a stand-alone bound-state by a
perturbative NLO correction. For each finite $\hbar \omega $ adopted in the
HO-wavefunction based calculations, one can always adjust $c_{nlo}$ and $%
d_{nlo}$ so that $E_{x}^{NLO}+E_{x}^{LO}$ is negative, but its nearest
state---which is either $E_{x+1}^{NLO}+E_{x+1}^{LO}$ or $%
E_{x-1}^{NLO}+E_{x-1}^{LO}$---is positive. This leads to the following
equations 
\begin{align}
c_{nlo}C_x+d_{nlo}D_x& =X,  \nonumber \\
c_{nlo}C_{x-1}+d_{nlo}D_{x-1}& =Y,  \label{eqsi}
\end{align}%
where I have assumed $x-1$ is the
nearest spectrum to $x$ (if not, $C_{x-1}$, $D_{x-1}$ need to be changed to $C_{x+1}$, $D_{x+1}$). And 
\begin{align}
X& =-(E_{x}^{LO}+\sigma ),  \nonumber \\
Y& =-(E_{x-1}^{LO}-\gamma ),  \label{deqs}
\end{align}%
with $\sigma ,\gamma >0$ so that $E_{x}^{NLO}+E_{x}^{LO}$ is negative and $%
E_{x-1}^{NLO}+E_{x-1}^{LO}$ is positive. Eq.~(\ref{eqsi}) leads to 
\begin{align}
c_{nlo}& =\frac{D_{x-1}X/D_x-Y}{D_{x-1}C_x/D_x-C_{x-1}},  \nonumber \\
d_{nlo}& =\frac{C_{x-1}X/C_x-Y}{C_{x-1}D_x/C_x-D_{x-1}}.  \label{sol}
\end{align}%
Now, I investigate the behavior of $c_{nlo}$ and $d_{nlo}$ in the limit $%
\hbar \omega \rightarrow 0$. In this limit, the spectra become continuous,
and the difference between $C_{i}$ and $C_{i-1}$ and $D_{i}$ and $D_{i-1}$ becomes
infinitesimal. Thus, they are related by an infinitesimal number $\epsilon >0
$: 
\begin{align}
C_{x-1}& =C_x-\epsilon ,  \nonumber \\
D_{x-1}& =D_x-\epsilon .  \label{eps}
\end{align}%
On the other hand, for a single bound-state to exist, the energies $X$ and $Y
$ must differ by a finite number. Let $\Delta =\sigma +\gamma $ in Eq.~(\ref%
{deqs}), one then has 
\begin{equation}
Y=X+\Delta ,  \label{delta}
\end{equation}%
in the limit $\hbar \omega \rightarrow 0$, where $\Delta >0$ is a finite
number. Substituting Eq.~(\ref{eps}) and Eq.~(\ref{delta}) into Eq.~(\ref%
{sol}), one obtains 
\begin{align}
c_{nlo}& =-\frac{\Delta }{\epsilon },  \nonumber   \\
d_{nlo}& =-\frac{\Delta }{\epsilon }. \label{limit}
\end{align}%
The above solution blows up in the limit $\hbar \omega \rightarrow 0$. Thus,
similar to what happened in case I, unless $E_{R}\rightarrow 0$ or $\Gamma
\rightarrow 0$, it is impossible to generate a stand-alone bound-state under
first-order perturbation theory---a striking difference to the
non-perturbative treatment. 

Now one can verify the above conclusion via
numerical calculations. I keep $c_{lo}=-2.5\times 10^{-9}$ MeV$^{-4}$ 
and vary $c_{nlo}=2\times 10^{-9}$ to $7\times 10^{-9}$ MeV$^{-4}$, $%
d_{nlo}=-1\times 10^{-13}$ to $-6\times 10^{-13}$ MeV$^{-6}$.
As one can see in Fig.~\ref{3p0_cd}, cascades of negative-energy states appear under the perturbative treatment. 
In fact, those states are only
separated due to the infrared truncation (as a finite $\hbar \omega $ is adopted). This can
be inferred from the decrease of binding energies and increase of the number
of negative-energy states as one approaches the continuum limit ($\hbar \omega
\rightarrow 0$, $n_{max}\rightarrow \infty $). On the other hand, when $%
V_{LO}+V_{NLO}$ are treated non-perturbatively, a single and deeper bound-state is created within
the range $c_{nlo}$ and $d_{nlo}$ tested, and binding energies stay
invariant under different $\hbar \omega $.
\onecolumngrid\
\begin{center}\
\begin{figure}[h]\
\includegraphics[width=\linewidth]{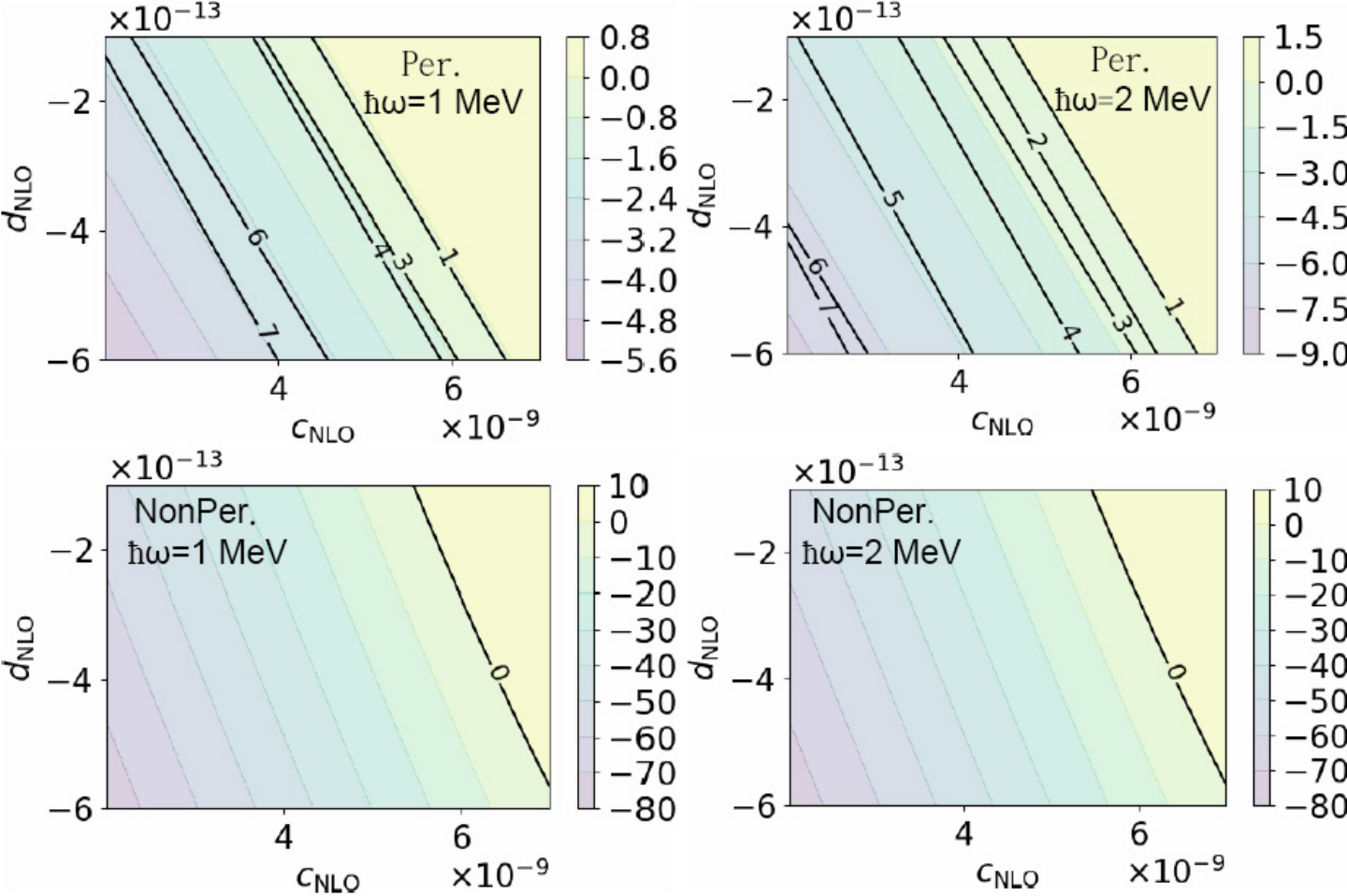}\
\caption{(Upper panels) The lowest state of $E^{LO}+E^{NLO}$ under
various $c_{nlo}$ and $d_{nlo}$, where the number of bound-states are given within the contour lines.
(Lower panels) Same as the upper panels, but the NLO interaction are
treated non-perturbatively.}\
\label{3p0_cd}\
\end{figure}\
\end{center}\
\twocolumngrid\

\section{Virtual states to bound-states}

\label{1s0} Virtual states are poles in the (analytic continued) scattering
amplitude which occur at momentum $k=i\gamma$ ($\gamma < 0$), with the
corresponding energies located in the second (unphysical) Riemann sheet~\cite%
{Burke2011}. It is well-known that a shallow virtual state exists in the NN $%
^1$S$_0$ channel, which has been formulated analytically within the pionless
EFT already at LO~\cite{ksw,ksw1,vanKolck:1998bw,1s0d}. For example, in
standard pionless EFT, one has~\cite{vanKolck:1998bw} 
\begin{align}
T_{LO}(k,k;k^2/M)&\approx\frac{4\pi}{M(1/a+ik-rk^2/2+...)}  \nonumber \\
&\approx\frac{-1}{1/c_{lo}-I_0(k)+...},  \label{vir}
\end{align}
where $a$ and $r$ are the scattering length and effective range, ``..."
stands for higher-order contribution, and 
\begin{align}
I_0(k)=-\frac{M}{2\pi^2}[\theta\Lambda +i\frac{\pi}{2}k+\frac{k^2}{\Lambda}O(%
\frac{k^2}{\Lambda^2})],
\end{align}
where $\theta$ is a positive number, and $\theta\Lambda=\Lambda$ if a sharp
cutoff $\Lambda$ replaces the regulator listed in Eq.~(\ref{r}). Thus, as
long as $a<0$ and $r>0$, the pole of the above T-matrix occurs at $k=i\gamma$
with $\gamma < 0$, which then corresponds to a virtual state. When $%
a\rightarrow -\infty$, this virtual state moves to the threshold.

\begin{figure}[tbp]
\includegraphics[scale=0.3]{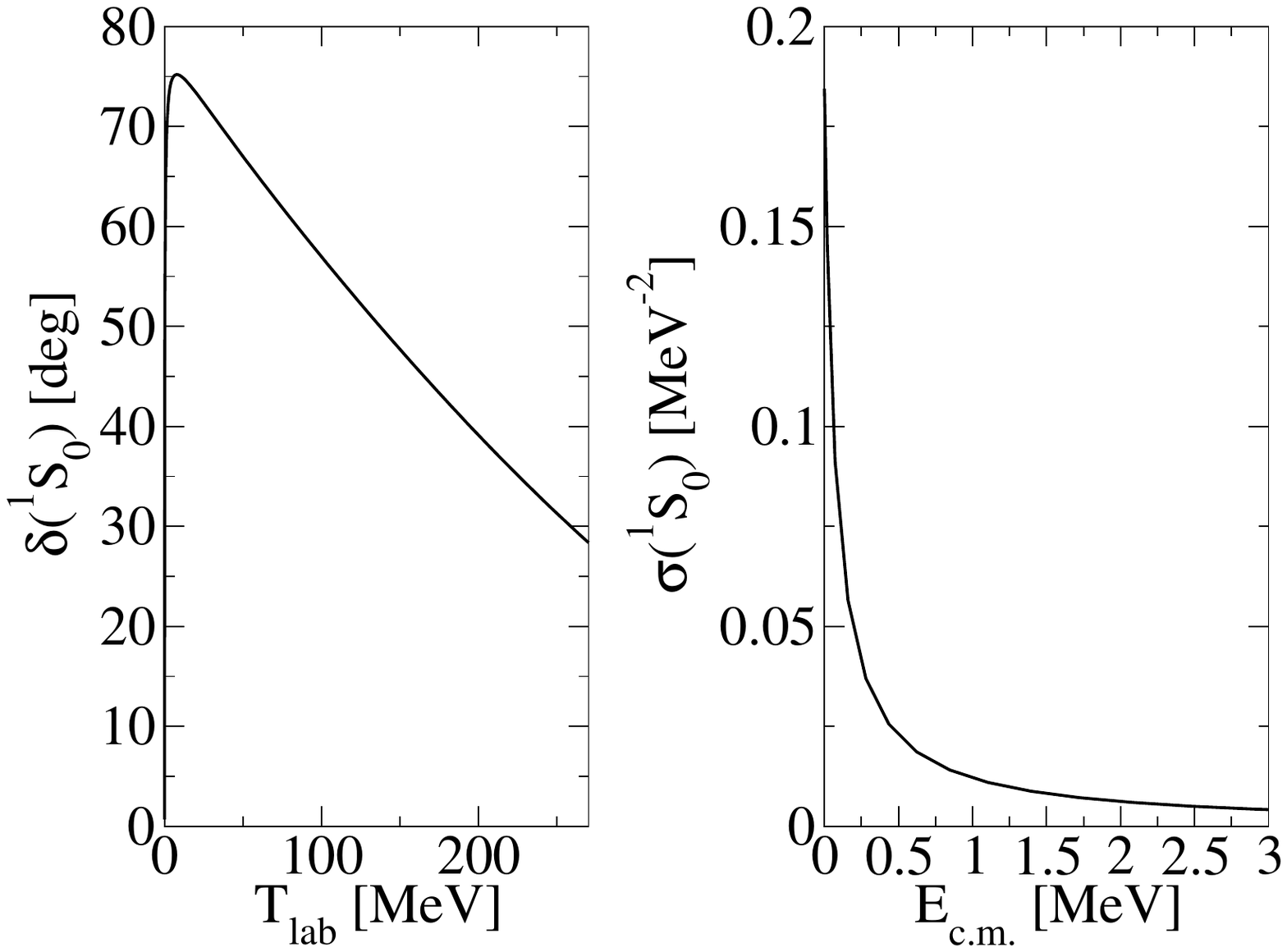}
\caption{Phase shifts $\protect\delta$ and cross section $\protect\sigma$ as
a function of laboratory energy $T_{lab}=2E_{c.m.}$ at LO for $%
c_{lo}=-4.7\times 10^{-6}$ MeV$^{-2}$ and $\Lambda=450$ MeV.}
\label{1s0_phase}
\end{figure}

In the following, I use NN $^1$S$_0$ scattering as a testing ground and
adopt 
\begin{align}
V_{LO}(p,p')&=c_{lo}f_R,  \nonumber \\
V_{NLO}(p,p')&=[c_{nlo}+d_{nlo}(p^{2}+p^{\prime}{}^2)]f_R,  \label{vs}
\end{align}
to investigate the possibility of converting a virtual state to bound-state
perturbatively. At LO, I adopt $c_{lo}=-4.7\times 10^{-6}$ MeV$^{-2}$ and 
$f_R$ as listed in Eq.~(\ref{r}) with $\Lambda=450$ MeV. Solving this $V_{LO}$ in
LSE gives $a=-24$ fm, $r=0.81$ fm, with the corresponding phase shifts and
cross section plotted in Fig.~\ref{1s0_phase}.

\begin{figure}[tbp]
\includegraphics[scale=0.3]{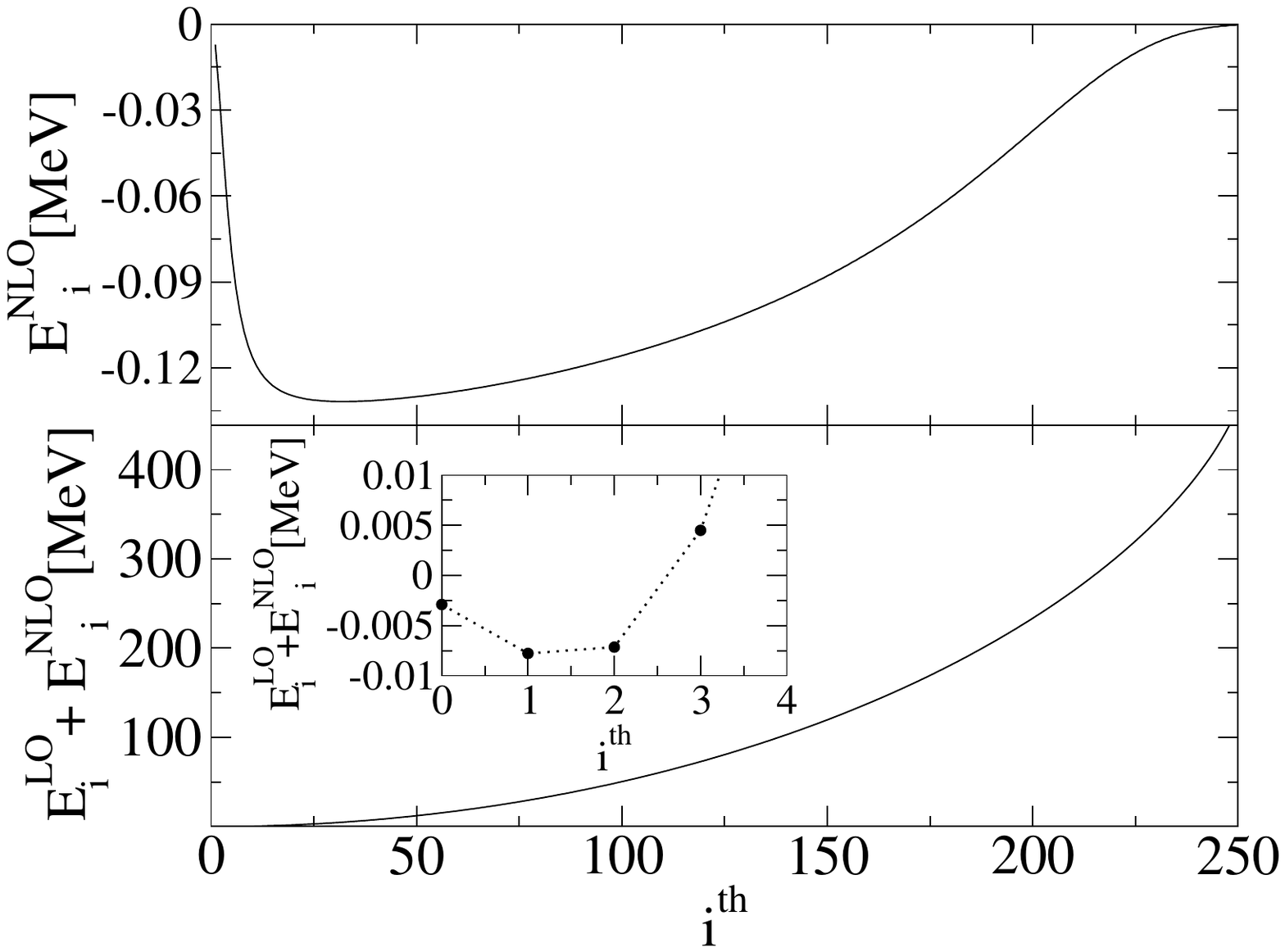}
\caption{$E^{NLO}_i$ (upper panel) and $E^{LO}_i+E^{NLO}_i$ (lower panel)
under $c_{lo}=-4.7\times 10^{-6}$ MeV$^{-2}$ $c_{nlo}=-20\times 10^{-7}$ MeV$%
^{-2}$ and $\Lambda=450$ MeV. The first four energies of the lower panel
are magnified in the inset. Note that $\hbar\protect\omega=1$ MeV is
adopted, therefore all the energies are discretized.}
\label{1s0_nlo}
\end{figure}

To investigate the NLO contribution, I again adopt $\hbar \omega =1$
MeV and evaluate Eq.~(\ref{enlo}). With $d_{nlo}=0$ and $c_{nlo}$ made
increasingly attractive, results analog to Table~\ref{t1} are listed in
Table~\ref{t1s}. One can see that Table~\ref{t1s} resembles what happened in
the $^{3}$P$_{0}$ channel, i.e., a virtual state is converted to a series of
bound-states as $V_{NLO}$ is made more attractive. $E^{NLO}_i$ and the total energy up to NLO
correspond to the most attractive case in Table~\ref{t1s} are plotted as Fig.~\ref{1s0_nlo}.
Once again, the NLO correction peaked around the index $x$ where $E^{LO}_x$ is the closest to
the energy of the virtual state. 

\begin{table}[h]
\begin{tabular}{ccc}
\hline\hline
$c_{nlo} $ (MeV$^{-2}$)  & B.S.$_{per}$ ($10^{-3}$ MeV) & B.S.$%
_{non-per}$ (MeV) \\ \hline
$-1\times 10^{-7}$  & --- & --- \\ 
$-2\times 10^{-7}$  & --- & $-7.5\times 10^{-4}$ \\ 
$-5\times 10^{-7}$  & --- & -0.21 \\ 
$-10\times 10^{-7}$ & --- & -1.2 \\ 
$-15\times 10^{-7}$ & -1.13, -1.61 & -2.9 \\ 
$-20\times 10^{-7}$  & -2.9, -7.8, -7.2 & -5.1 \\ \hline\hline
\end{tabular}%
\caption{NLO results in $^1$S$_0$ channel under $c_{lo}=-4.7\times 10^{-6}$
MeV$^{-2}$ and $\Lambda=450$ MeV. Results are obtained via HO-wavefunctions
with $\hbar\protect\omega=1$ MeV and $n_{max}=250$. 
B.S.$_{per}$ (B.S.$_{non-per}$) is the bound-state energy obtained at NLO
when treating Eq.~(\protect\ref{r1}) perturbatively (non-perturbatively). ``---" denotes that no bound-state is generated. }
\label{t1s}
\end{table}

Next, I consider both $c_{nlo}$ and $d_{nlo}$. I kept $V_{LO}$ the same
and vary $c_{nlo}=1\times 10^{-6}$ to $5\times 10^{-5}$ MeV$^{-2}$, $%
d_{nlo}=-2\times 10^{-11}$ to $-5.2\times 10^{-10}$ MeV$^{-4}$. Results are
presented in Fig.~\ref{1s0_cd}. As one can see, cascades of bound-states appear in a way similar to the NN $^3$P$_0$ case. 
On the other hand, when $%
V_{LO}+V_{NLO}$ are treated non-perturbatively, a single and deeper bound-state is created and the binding energies stay
invariant under different $\hbar \omega $.
\onecolumngrid\
\begin{center}\
\begin{figure}[h]\
\includegraphics[width=\linewidth]{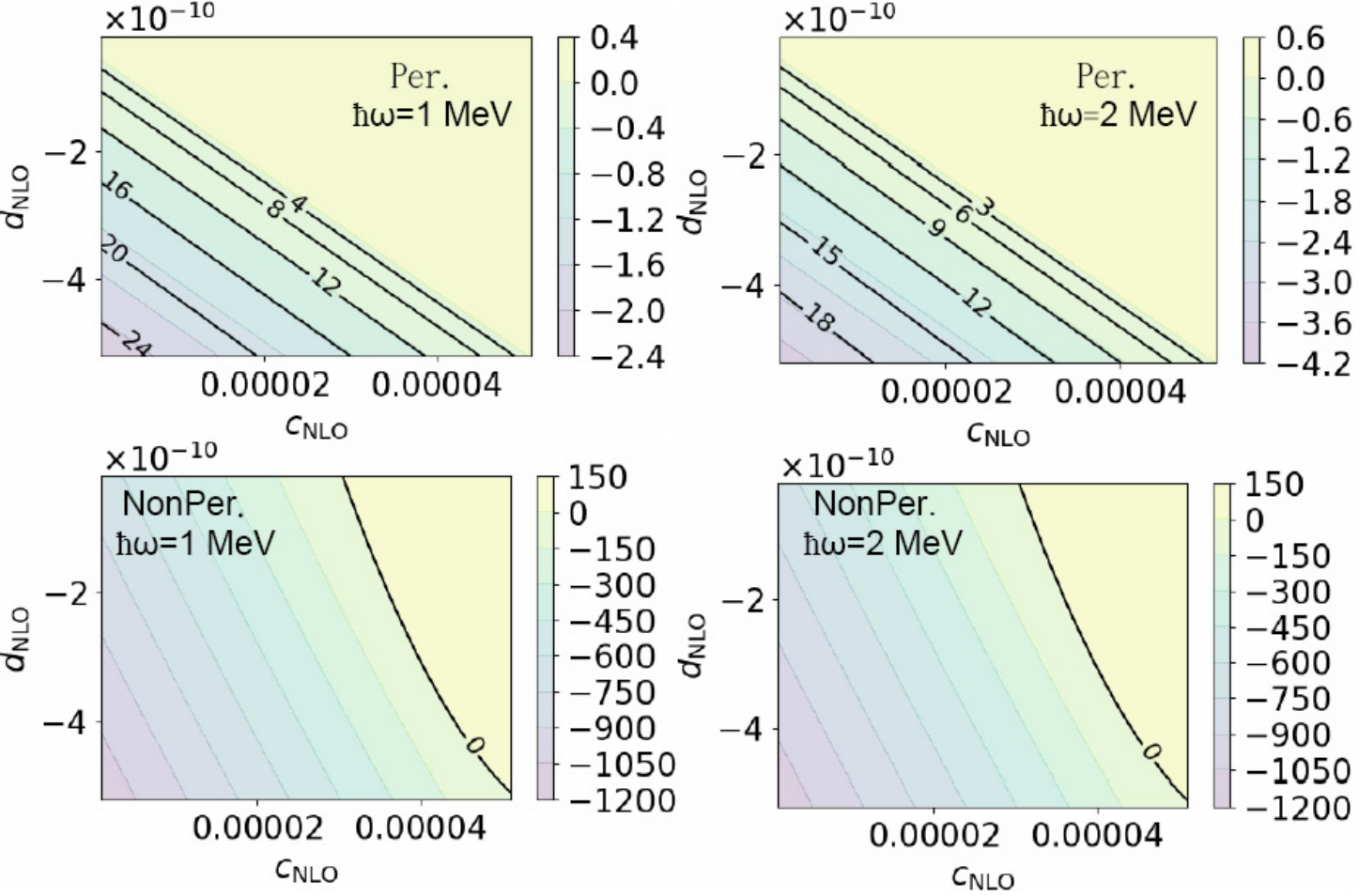}\
\caption{(Upper panels) The lowest state of $E^{LO}+E^{NLO}$ under
various $c_{nlo}$ and $d_{nlo}$, where the number of bound-states are given within the contour lines.
(Lower panels) Same as the upper panels, but the NLO interaction are
treated non-perturbatively.}\
\label{1s0_cd}\
\end{figure}\
\end{center}\
\twocolumngrid\

\section{A reverse case}
\label{rev}
Considering now a scenario where a bound-state is generated from a non-perturbative treatment of the LO interaction, and one 
wishes to remove it at NLO perturbatively. A repulsive $V_{NLO}$ is then required in this case, and its contribution is again given by Eq.~(\ref{enlo}).
Only now the eigen-function corresponds to the LO bound-state can be easily represented by HO-wavefunctions as they are of the same type. 
In fact, since the LO bound-state wavefunction and $V_{NLO}$ have more overlap within finite distance, one can easily design a scenario
that the required NLO energy-shift to lift the LO bound-state are small so that the corresponding $V_{NLO}$ required is
much smaller than $V_{LO}$. 
Thus, everything appear to be well-defined within perturbation theory.

However, there is still a problem. The bound-state lifted from LO will now join the continuum at NLO. Thus, there will be two states correspond to one energy in the NLO
continuum spectrum---one belongs to a state which is shifted upward from the LO continuum, the other is the LO bound-state which contains a pole.
In general, with the presence of resonances or virtual states, level-crossing within the continuum spectrum can happen from LO to NLO under perturbation theory, 
where one can always find two different states coincide at NLO as they can receive 
unequal energy shifts with the difference equals to their LO gap. 
This degeneracy is not a problem as long as their wavefunctions evaluated up to NLO only differ by higher-order components.    
However, the coincidence of the lifted bound-state to another continuum state poses several conceptual difficulties. 
First, it is not entirely clear whether the NLO wavefunction of the lifted bound-state is in presence of bound or continuum property.
Moreover, the T-matrix at this energy is degenerated with a difference much more than what higher-orders in perturbation theory can account for. 
I.e., one is finite and the other has a pole. Note that this pole presences a particular feature, as in its nearby energy region, the T-matrix does not
have the trend to become divergent at all. Thus, this state must be categorized as an unphysical state and be discarded, even though the shift from LO to NLO
can be completely legal within perturbation theory. 
On the other hand, poles of this type are exactly what one needed in the previous two sections in order to create stand-alone bound-states perturbatively.  
One cannot claim that this type of poles is the \textit{only} one to be used to create problem-free bound-states perturbatively. 
Unfortunately, this is the most straightforward and probably the only scenario which is numerically possible for $A>4$ systems.

\section{A numerical example of lifting the leading order bound-state}
\label{rev_num}

\begin{figure}[tbp]
\includegraphics[scale=0.3]{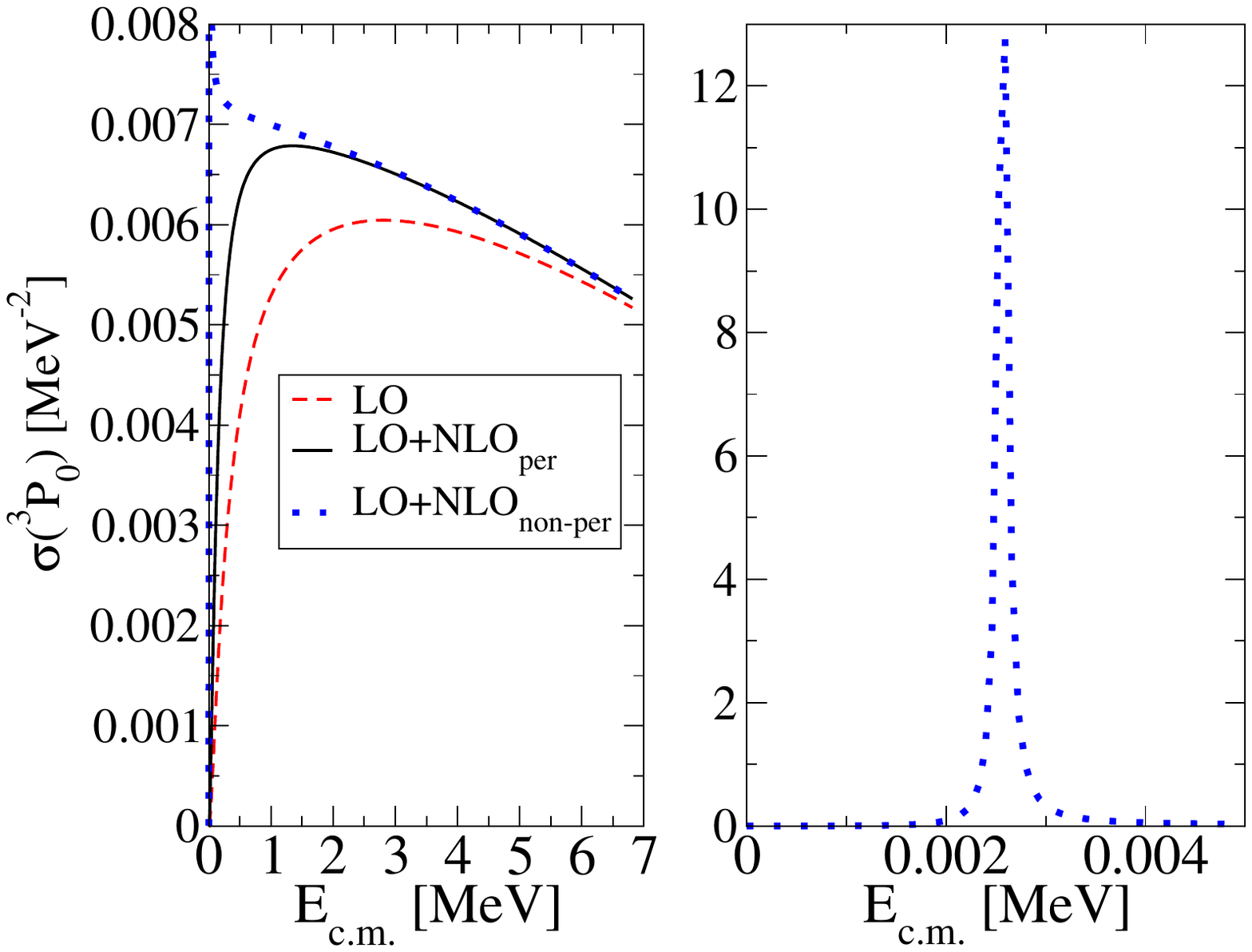}
\caption{Cross section $\protect\sigma$ as a function of c.m. energy $E_{c.m.}$ corresponds to the last row of
Table~\protect\ref{rev_t1} (i.e., with $c_{nlo}=8\times 10^{-11}$ MeV$^{-4}$), where the LO+NLO$_{non-per}$ results (blue dots) are re-plotted in the right panel in order to
show its full amplitude around the peak.}
\label{rev_cross}
\end{figure}

I now present an example regarding the removal of LO bound-states as postulated in Sec. \ref{rev}. Under the NN $^3$P$_0$ case and $\Lambda=150$ MeV in Eq.~(\ref{r}), adopting
$V_{LO}=c_{lo}pp'$ with $c_{lo}=-2.8\times 10^{-9}$ MeV$^{-4}$ will gives a LO bound-state with binding energy $E^{LO}_0=-0.194$ MeV.
Then, I add a $V_{NLO}$ which has the same structure as $V_{LO}$ but is 35 times weaker, i.e., $c_{nlo}=8\times 10^{-11}$ MeV$^{-4}$ and $d_{nlo}=0$ as listed in Eq.~(\ref{r1}). 
In this case, the LO bound-state receives a perturbative lift $\approx 0.211$ MeV and join the continuum at NLO.  
Note that if the same $c_{nlo}$ is added non-perturbatively, the
correction from LO to NLO is $0.197$ MeV.
In Table~\ref{rev_t1}, the
perturbative and non-perturbative results are compared with $c_{nlo}$ ranging from $5\times 10^{-11}$ to $8\times 10^{-11}$ MeV$^{-4}$.
As one can see, at least up to $8\times 10^{-11}$ MeV$^{-4}$, the energy shift received by the LO bound-state can be considered small, as
the non-perturbative results can be approximated perturbatively with an error well-below $10\%$. 
\begin{table}[h]
\begin{tabular}{ccc}
\hline\hline
$c_{nlo} $ (MeV$^{-4}$)  & $E^{NLO}_0$ (MeV) & $E^{NLO}_{non-per,0}$ (MeV) \\ \hline
$5\times 10^{-11}$ & 0.132 & 0.127 \\ 
$6\times 10^{-11}$  & 0.159 & 0.151 \\ 
$7\times 10^{-11}$& 0.185 & 0.175 \\ 
$8\times 10^{-11}$  & 0.211 & 0.197 \\ \hline\hline
\end{tabular}%
\caption{The perturbative versus non-perturbative NLO shifts to the LO bound-state, where results are obtained 
in the NN $^3$P$_0$ channel under $c_{lo}=-2.8\times 10^{-9}$ MeV$^{-4}$ and $c_{nlo}$ with values listed above.
Calculations here converge within $0.1\%$ with the decrease of $\hbar\omega$ and increase of $n_{max}$.
}
\label{rev_t1}
\end{table}
However, cross sections obtained perturbatively and non-perturbatively up to NLO show another story, where a big difference between 
the two NLO treatments is presented at $E_{c.m.}\approx 0.0026$ MeV as shown in Fig.~\ref{rev_cross}.  
Note that the perturbative amplitude up to NLO are generated from $T_{LO}(p_0,p_0;E_{c.m.})$ with $E_{c.m.}>0$, so the NLO results
do not contain the state converted from the LO pole\footnote{The state corresponds to the lifted-pole needed to be generated by inserting 
$T_{LO}(p_0,p_0;E_{c.m.})$ with $E_{c.m.}=-0.194$ MeV into
Eq.~(\ref{eqn:LSE23}), which is however not directly calculable since
$T_{LO}$ diverges.}.
As a result, the perturbative generated cross section is smooth against $E_{c.m.}$ without any 
particular feature presented around the energy of the lifted pole $E_{c.m.}=E^{LO}_0+E^{NLO}_0\approx 0.017$ MeV.   
On the other hand, a sharp resonance is generated via the non-perturbative treatment with its full shape around the peak re-plotted in the right panel of Fig.~\ref{rev_cross}---where the energy at the peak of the resonance $\approx 0.0026$ MeV can be understood as a cancellation between the LO eigen-energy (-0.194 MeV) and the inferred NLO non-perturbative correction (0.197 MeV). 
The difference between the two NLO treatments are rather small after $E_{c.m.}>1$ MeV. Moreover, the change from LO to NLO is $\lessapprox 15\%$, which
suggests that perturbation theory works very well in this region.
However, the mis-match between results obtained perturbatively and non-perturbatively at $E_{c.m.}<1$ MeV suggests that
 even one can make a case where the NLO interaction 
is arbitrarily small compared to $V_{LO}$, as long as it involves a sign change in the pole position,
there will be an energy domain where the observables obtained from these two approaches disagree.  

The lesson one learned from the above case can be summarized as the follows. If an energy-shift from LO to NLO involves bound to/from unbound transitions, but 
the observables one wishes to describe do not involve the problematic part of this conversion, then the perturbative corrections will agree with the non-perturbative results to a degree similar to cases where there is no threshold crossing. 
Meanwhile, there is always an energy domain where non-perturbative and perturbative threshold-crossing disagree with each other. 
%

\section{Essence of the problem}
\label{avoid}
Results in the above sections show that, although a transition between bound and unbound states can be achieved
by perturbative corrections, the outcome is not necessarily desirable. 

Since any unbound to/from bound transition needs to go through the threshold, the problem can be
reduced into whether the threshold can be (or how well it can be) produced perturbatively.
In the following I consider the simplest case, i.e., a system consists of only two particles.  
Note that scattering states are continuous in $E$. On the other hand, for $E<0$, there are discretized bound-states. Now, consider the situation where
the interaction is made just attractive enough so that an $E=-|\epsilon|$ state is produced.   
When this is produced non-perturbatively, it presents an unique and universal feature---i.e., the so-called unitarity feature~\cite{Hammer:2006ct,Braaten:2004rn,Griesshammer:2023scn,vanKolck:2019qea,vanKolck:2018vzl,vanKolck:2017jon,Konig:2016utl}. The scattering length is infinity at threshold.
Note that due to the avoided-crossing phenomenon, no continuous bound-state can be generated.
On the other hand, several problems could occur when the threshold-crossing needed to be 
executed perturbatively.
First,
there is no guarantee that when the real part of an eigen-energy $Re(E)=-|\epsilon|$ ($\epsilon\rightarrow0$) is first generated, the scattering length is still much larger than other scales.
This can be understood by rewriting the NLO potential
to be evaluated within the first-order perturbation theory as $\chi v_{nlo}$, where 
$\chi$ is an adjustable number denoting the overall strength and $v_{nlo}$ concerns the form of the 
interaction. Various combinations of $\chi$ and $v_{nlo}$ can give the same real number $\langle\Psi_{i}^{LO} |\chi v_{nlo}|\Psi_{i}^{LO}\rangle$---which cancels the real part of the LO eigen-energy---so that at least the real-axis satisfies the threshold condition.
However, this freedom also means that at least one more scale can enter (e.g., various values of scattering length and effective range can be accommodated by adjusting $\chi$ and the form of $v_{nlo}$ while keeping $\langle\Psi_{i}^{LO} |\chi v_{nlo}|\Psi_{i}^{LO}\rangle$ constant), and the unique feature of approximate scale invariance near the unitarity limit does not hold in general if the threshold is approached solely via the first-order perturbation theory. 
 Since the unitarity of the S-matrix is no longer exact under the perturbative treatment, any physical observable evaluated perturbatively within DWBA will in general come with an imaginary component. Normally this
does not pose a problem, as if the perturbation is justified, the non-unitary part will be small compared to the full results. 
However, when a bound-state is just created after the occurrence of threshold-crossing, any non-vanishing imaginary component in $E$ will be much larger compared to $|\epsilon|$, so that the unitarity of S-matrix is severely violated.
Since the Riemann sheet where resonances reside is connected to the one where bound-states belong to only through the origin, any defect with respect to the exact threshold will result in the presence of a non-zero width of the bound-states and being magnified linearly with the growing of $\chi$, leading to the phenomenon as seen in the previous sections. 

Thus, unless one chooses a very specific $v_{nlo}$ and/or applies additional procedures
to ensure that the unitarity is well-reproduced at threshold, the two approaches (perturbative v.s. non-perturbative) will generate 
results with very different structure---even within the energy domain where the corrections are
small in both cases.

\section{Summary and implications to many-body systems}
\label{summary}
In this work the possibility of generating bound-states perturbatively has been tested. 
Although numerical calculations are performed only on several selective examples, the results can be
very general and will hold as long as: (i) the interactions are finite-ranged; (ii) the problem can be evaluated under 
the time-independent Schr\"odinger equation---where all eigen-values are real numbers.

The investigation suggests the following:

\begin{itemize}
\item Under the existence of resonances or virtual states---which are
generated non-perturbatively at LO---corrections in energy due to a
perturbative insertion of NLO interaction peak at the resonance energy 
or the momentum corresponds to the pole position of the virtual
state. Therefore, re-ordering of the original eigen-states happens, and since
level-crossing is allowed, it is possible to convert resonances and virtual states to negative-energy states
perturbatively.

\item However, in contrast to the non-perturbative treatment, a straightforward implementation of DWBA will generally result in a series of perturbatively created negative-energy states (or to be interpreted as a bound-state with non-zero width).

\item On the other hand, an attempt to remove a LO bound-state perturbatively also leads to disagreements with respect to the non-perturbative results, at least in some particular energy domain.

\item Therefore, moving a pole perturbatively across the threshold (which corresponds to a transition between the unphysical and physical sheets) requires extra care. In general, a stand-alone ground-state cannot be produced by simply applying the first-order perturbation theory.

\end{itemize}

The study in two-particle systems suggests the following generalization for
an $A$-particle system. Suppose after solving $V_{LO}$ non-perturbatively,
an $A$-particle system consists of a bound sub-system with $(A-x)$ particles and $x$
free particles. Then, by adding $V_{NLO}$ perturbatively, one could easily
make the $(A-x)$ sub-system more bound (provided that $V_{NLO}$ is an $n$%
-body force with $n \leq A-x$). Therefore, it could appear that the total binding energy of the
system can be shifted to any desired value. However, a change in the pole structure---which
requires the adherence between the $(A-x)$ sub-system and $x$ free particles---is
unlikely to happen without creating a cascade of bound-states.

In summary, with the caveat that the perturbative correction examined in this work is just a subset of the perturbative EFT-power counting, I conclude that perturbative threshold-crossing is possible, but there is always a region of energy within which it gives results very different from those obtained non-perturbatively. Other than that energy domain, observables under the rest of the energy seem to follow the same rule governed by the common expectation from perturbation theory and behave similarly to those cases where there is no threshold-crossing. Meanwhile, this also means if the key observable one wishes to recover involves (or is passing through) the energy domain where a non-perturbative treatment is essential---e.g., the generation of bound-states---then a discretized spectrum of bound-states is unlikely to be generated perturbatively, even with the presence of resonances or virtual states at the
previous order. In this case, a resummation carefully guided by EFT or alternatives such as the ``improved action"~\cite{Symanzik:1983dc,Symanzik:1983gh,Lee:2008fa,Contessi:2023yoz} seems to be unavoidable.
%

%


\textit{Acknowledgements.} I thank D. Gazda for the assistance during various stages of this work, J. Mares, T. Dytrych and D. Phillips for useful discussions.
This work is partially stimulated from the (on-line) discussion of INT workshop---Nuclear Forces for Precision Nuclear Physics (INT-21-1b).   
This work was supported by the
the Extreme Light Infrastructure Nuclear Physics (ELI-NP) Phase II, a project co-financed by the Romanian Government and the European Union through the European Regional Development Fund - the Competitiveness Operational Programme (1/07.07.2016, COP, ID 1334);  the Romanian Ministry of Research and Innovation: PN23210105 (Phase 2, the Program Nucleu); the ELI-RO grant Proiectul ELI12/16.10.2020 of the Romanian Government; and the Czech Science Foundation GACR grant 19-19640S and 22-14497S. I acknowledge PRACE for awarding us access to Karolina at IT4Innovations, Czechia under project number EHPC-BEN-2023B05-023 (DD-23-83 and DD-23-157); IT4Innovations at Czech National Supercomputing Center under project number OPEN24-21 1892; Ministry of Education, Youth and Sports of the Czech Republic through the e-INFRA CZ (ID:90140) and CINECA HPC access through PRACE-ICEI standard call 2022 (P.I. Paolo Tomassini).
\bibliography{per_ref}

\section*{Appendix}

I compare here the results obtained by the HO-basis with the exact solution of LSE, in order to check whether the calculations presented in Sects.~\ref{3p0} and \ref{1s0} are reliable. The T-matrix as given in Eq.~(\ref{eq:2.3}) is analytically solvable if $V_{LO}$ has a simple form. Here, I consider the simplest case, i.e., $V_{LO}=c_{lo}$ in the $^1$ S$_0$ channel. With a sharp cutoff $\Lambda$, the integral kernel is
\begin{align}
&\frac{2}{\pi }M\int_{0}^{\Lambda }\frac{
dp^{\prime \prime }\;p^{\prime \prime }{}^{2}}{p_{0}^{2}+i\varepsilon -p^{\prime \prime }{}^{2}} \nonumber \\
&=\frac{2}{\pi }M\left[-\Lambda-i\frac{\pi}{2}p_0 +\frac{p_0}{2}\ln(\frac{\Lambda+p_0}{\Lambda-p_0})\right].
\label{a1}
\end{align}
It then follows that the corresponding $T_{LO}$ is independent of $(p,p')$ and is only a function of $E_{c.m.}=p_0^2/M$, i.e.,
\begin{equation}
T_{LO}=\frac{1}{\frac{1}{c_{lo}}+\frac{2M}{\pi}\left[\Lambda+i\frac{\pi}{2}p_0-\frac{p_0}{2}\ln(\frac{\Lambda+p_0}{\Lambda-p_0})\right]}.
\label{a2}    
\end{equation}
Note that the above expression differs from Ref.~\cite{vanKolck:1998bw} (as listed in Eq.~(\ref{vir})) due to a different convention in the plane-wave normalization and is reflected on the prefactor of the integral kernel.
The NLO contribution due to $V_{NLO}=c_{nlo}$ (setting $d_{nlo}=0$ in Eq.~(\ref{vs}) and replacing $f_R$ with a sharp cutoff $\Lambda$) is
\begin{equation}
T_{NLO}=c_{nlo}\left[1+F+\frac{F^2}{4} \right] ,
\label{a3}    
\end{equation}
with
\begin{equation}
F=\frac{4M}{\pi}T_{LO}\left[-\Lambda-i\frac{\pi}{2}p_0+\frac{p_0}{2}\ln(\frac{\Lambda+p_0}{\Lambda-p_0})\right].
\label{a4}    
\end{equation}
The phase shift is related to $T_{LO}$ and $T_{NLO}$ by Eq.~(\ref{lophase}) and Eq.~(\ref{nlophase}), respectively.

Note that further investigating the T-matrix for $E_{c.m.}<0$ requires an analytic continuation of Eq.~(\ref{a2})-(\ref{a4}) and involves expanding and reconciling the double pole structure within $F^2$, which will not be carried out in this work\footnote{In general, one needs to analytically continue the LSE. This is normally avoided by iterating $V_{LO}+V_{NLO}$ non-perturbatively at first, and then expand the T-matrix perturbatively~\cite{vanKolck:1998bw}.}. 

On the other hand, under HO-basis, one could convert eigenvalues and eigenfunctions to phase shifts through the J-matrix method~\cite{Yamani1975,Shirokov:2003kk,Yang:2016brl}.
It is shown~\cite{Yang:2016brl} that the converted phase shifts in general possess an oscillatory behavior, and approach the correct results (given by solving the LSE) when both the matrix element of the potential and the scattering wavefunction are captured/saturated by the infrared cutoff $\lambda\approx\sqrt{\frac{\hbar\omega M}{4(2n_{max}+7/2)}} $ and ultraviolet cutoff $\sqrt{M(2n_{max}+7/2)\hbar\omega}$ of the HO-basis~\cite{PhysRevC.89.044301,PhysRevC.90.064007}.

Note that when a sharp ultraviolet cutoff $\Lambda$ is applied, the particles interact up to $p_0=\Lambda$ and then the wavefunctions are forced to zero after the c.m. momentum $p>p_0$. This creates a problem in evaluating the scattering process via HO-basis. On one hand, the scattering wavefunction, i.e., Eq.~(\ref{10}), must be saturated by the HO wavefunction $\psi^{HO}_n$, with $n$ up to $n_{max}\gtrsim\frac{\Lambda^2}{2M\hbar\omega}$. On the other hand, as the increase of $n_{max}$, the orthogonality is no longer guaranteed if the momentum space integral is limited exactly at $p=\Lambda$, i.e., 
\begin{equation}
\int_0^{\Lambda}dpp^24\pi|\psi_{n\approx n_{max}}^{HO}(p)|^2\neq 1,
\label{a5}    
\end{equation}
which is un-desirable for the J-matrix method as part of its derivation relies on the orthogonality condition. 
In general, one avoids this dilemma by replacing the sharp cutoff $\Lambda$ by regulators, which then allows the potential to vanish smoothly and fulfills the saturation of ultraviolet physics and orthogonality at the same time. 

\begin{figure}[tbp]
\includegraphics[scale=0.33]{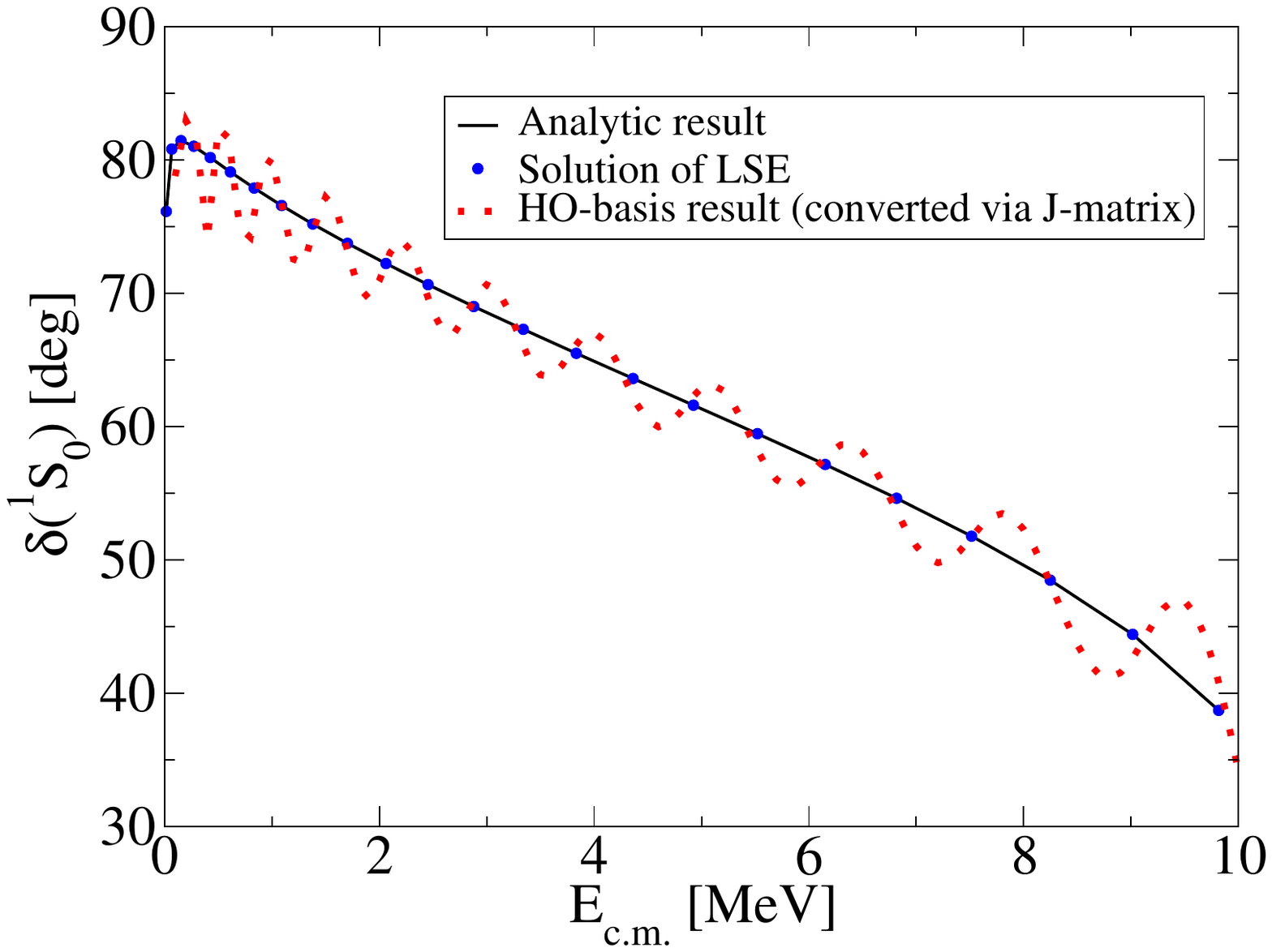}
\caption{Phase shifts as a function of the c.m. energy in the $^1$S$_0$ channel with $c_{lo}=-1.65\times 10^{-5}$ MeV$^{-2}$ at LO. The analytical results are obtained from Eq.~(\ref{a2}) and Eq.~(\ref{lophase}), which coincide with the numerical solution of LSE. The HO-basis results
are converted from the eigenvalues and eigenfunctions after diagonalizing $\langle H_{LO} \rangle$ with HO-wavefunctions $\hbar\omega=1$ MeV and $n_{max}=100$.}
\label{a1s}
\end{figure}

In the following, I will demonstrate the validity of describing scattering processes with the HO-basis chosen in this work. 
Although adopting a sharp cutoff has un-desirable features, it is still of interest to perform the first check based on it. In Fig.~\ref{a1s}, the simplest $^1$S$_0$ case as presented in Eq.~(\ref{a2}) with $\Lambda=100$ MeV and $c_{lo}=-1.65\times 10^{-5}$ MeV$^{-2}$ is considered (note that a slightly more attractive $c_{lo}=-1.68\times 10^{-5}$ MeV$^{-2}$ will produce a bound-state at LO). The phase shifts obtained analytically from Eq.~(\ref{a2}) are plotted as a function of $E_{c.m.}$ against the numerical solution of LSE (via methods as described in, e.g., Ref.~\cite{Yang:2007hb}) and the phase shifts converted by J-matrix method from the eigenvalues and eigenfunctions within the HO-space. As one can see, the numerical solution of LSE coincides with the analytic result, while the HO-basis result has an oscillatory behavior. Nevertheless, it is clear that, even under the unfavorable form of a sharp cutoff, the major feature of the scattering process can still be reasonably captured within the HO-space. Note that the HO-basis result presents the same oscillatory characteristic behavior for $n_{max}=14-100$, which does not die off with further increase of $n_{max}$ due to the effect of adopting a sharp cutoff. 

\begin{figure}[tbp]
\includegraphics[scale=0.33]{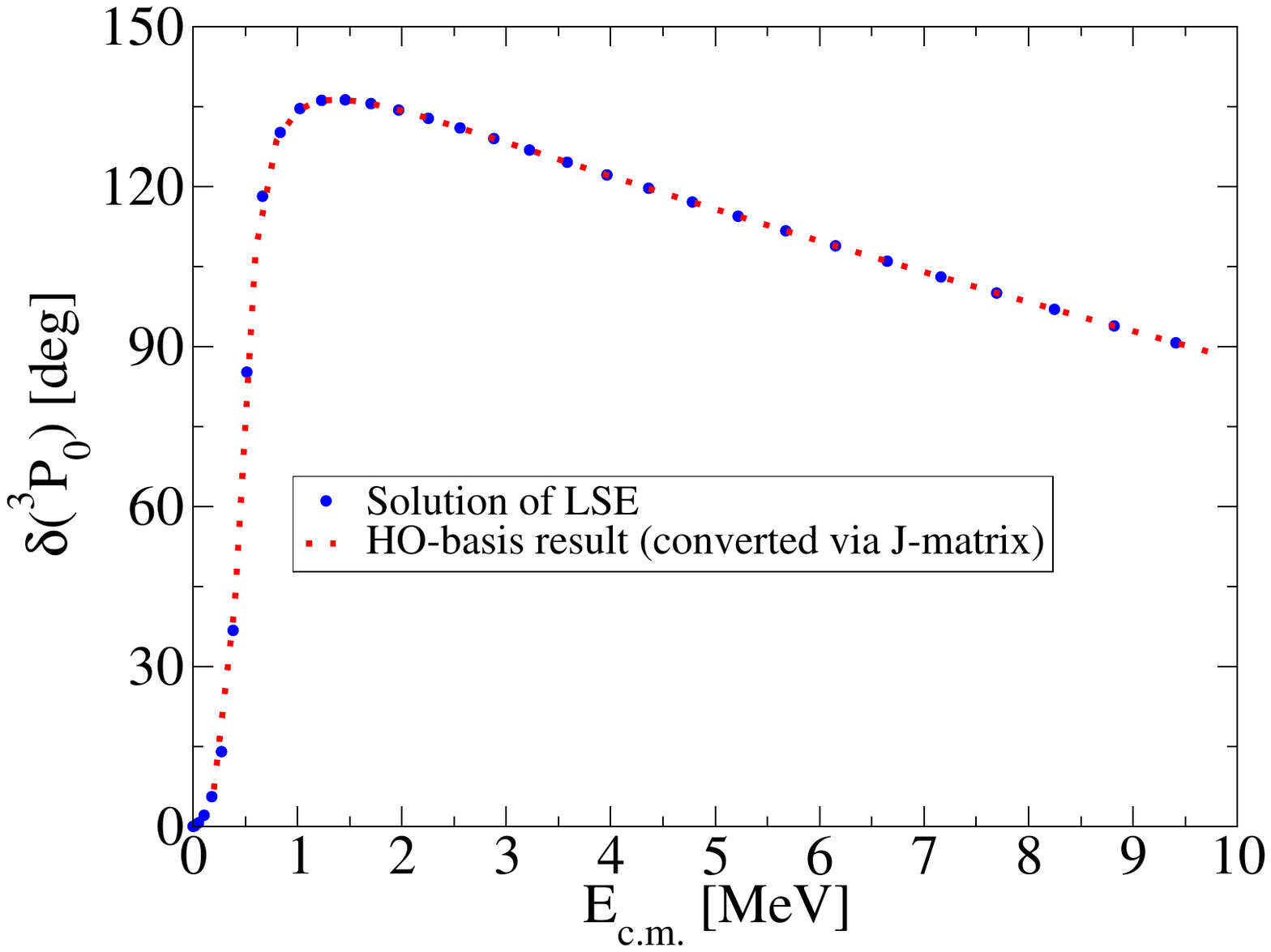}
\caption{Phase shifts as a function of the c.m. energy in the $^3$P$_0$ channel with $c_{lo}=-2.5\times10^{-9}$ MeV$^{-4}$. Here I compare the numerical solution of LSE and the HO-basis results
converted from the eigenvalues and eigenfunctions after diagonalizing $\langle H_{LO} \rangle$ with HO-wavefunctions $\hbar\omega=1$ MeV and $n_{max}=100$.}
\label{a3p}
\end{figure}

In the next check, the regulator $f_R$ as listed in Eq.~(\ref{r}) is adopted and a LO $^3$P$_0$ setup---which is identical to the one presented in Fig.~\ref{3p0_phase}---is compared with the results obtained in the HO-space (again converted via the J-matrix method) in Fig.~\ref{a3p}. With the regulator, the phase shifts given by the J-matrix method with $n_{max}=100$ reproduce the scattering process almost exactly.  

\begin{figure}[tbp]
\includegraphics[scale=0.33]{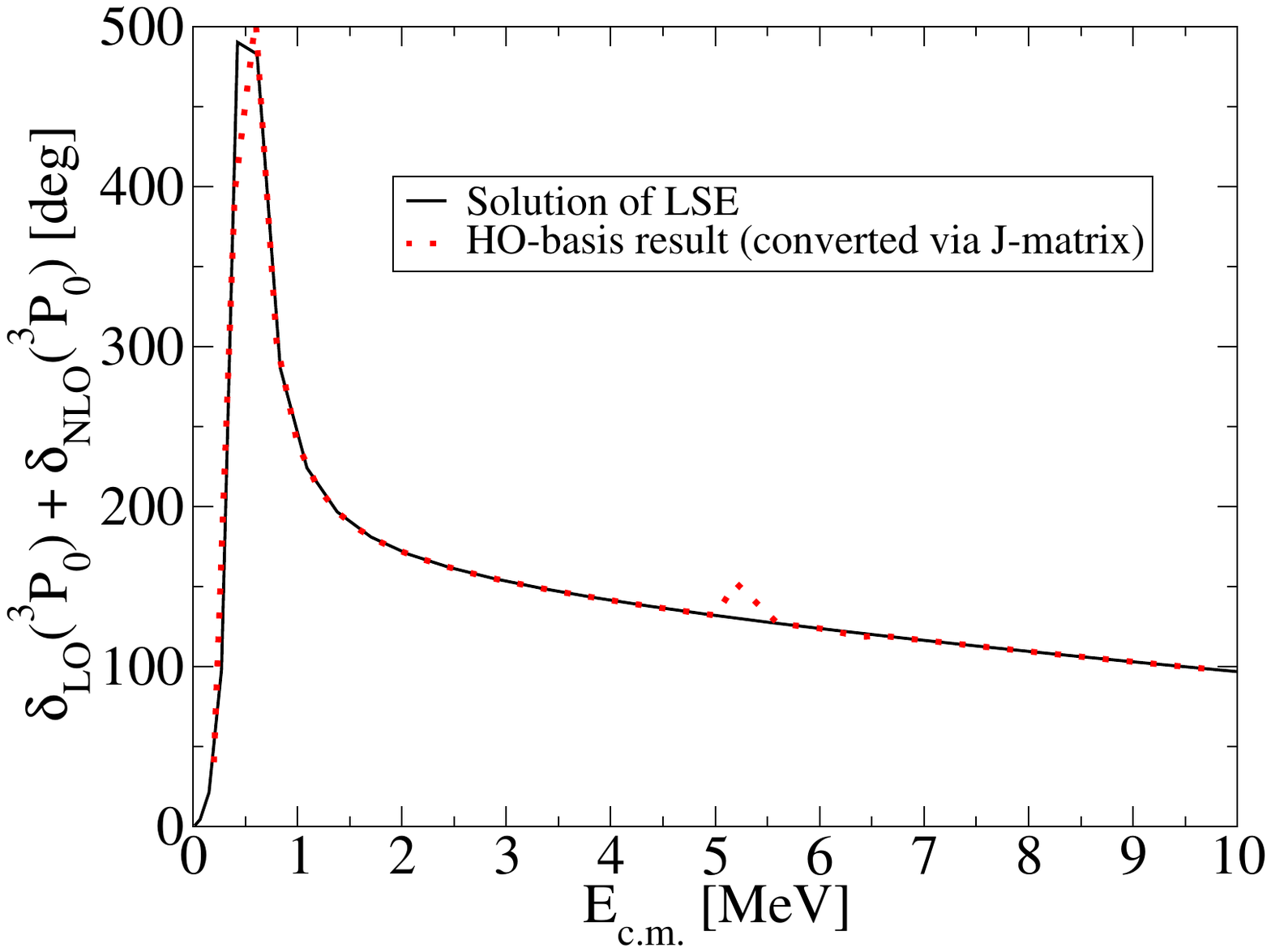}
\caption{Phase shifts up to NLO as a function of the c.m. energy in the $^3$P$_0$ channel, where the setup is identical to Table~\ref{t0}. The HO-basis results
are obtained by the J-matrix utilizing the Hellmann-Feynman theorem, with HO-wavefunctions $\hbar\omega=1$ MeV and $n_{max}=100$. The bump around $E_{c.m.}=5.2$ MeV is of numerical origin.}
\label{3p0_s}
\end{figure}

\begin{figure}[tbp]
\includegraphics[scale=0.33]{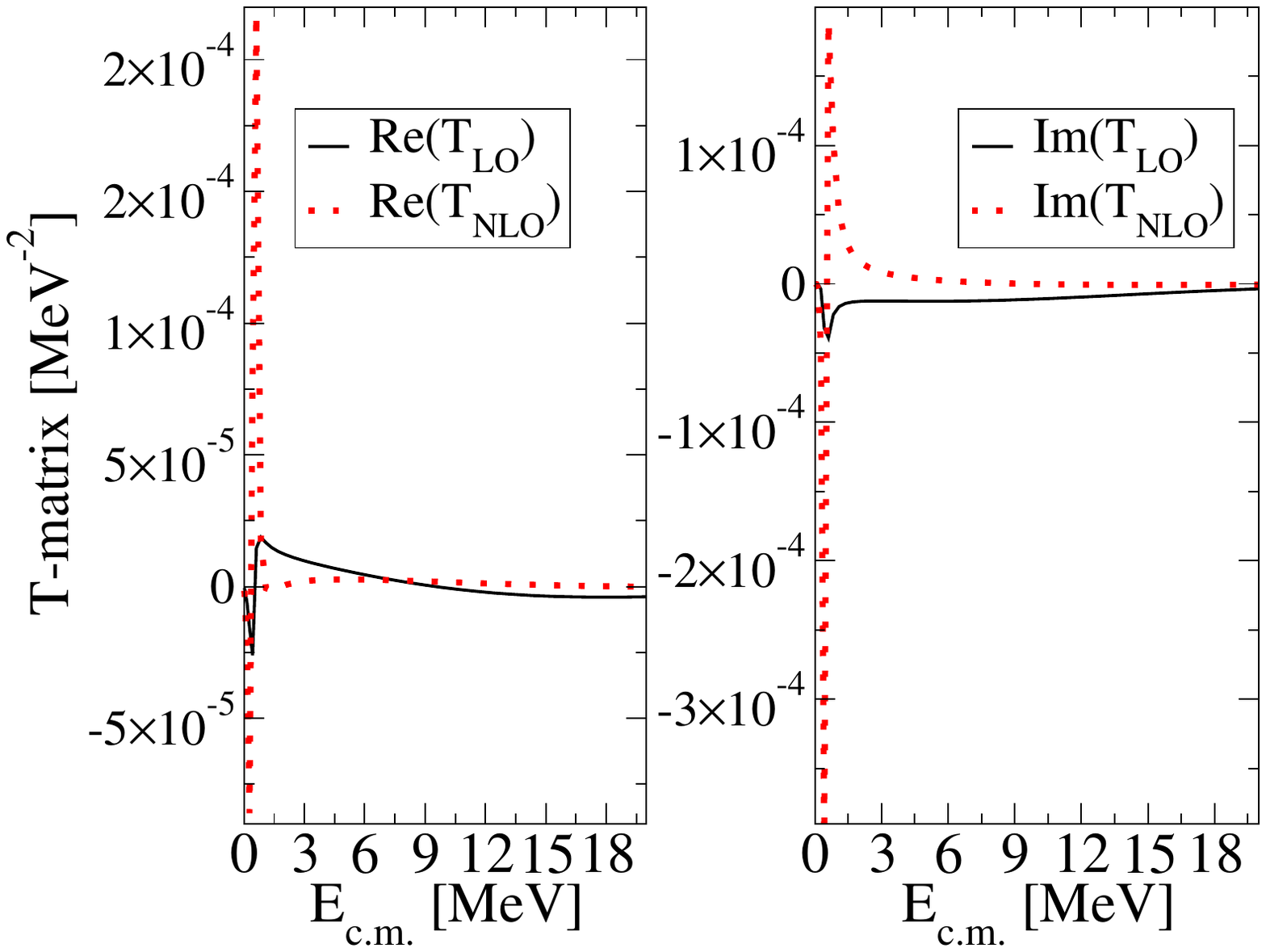}
\caption{The real (Re) and imaginary (Im) part of the LO and NLO T-matrix as a function of $E_{c.m.}$. Here $T_{LO}$ and $T_{NLO}$ are obtained using numerical solutions of the LSE and Eq.~(\ref{eqn:LSE23}).}
\label{3p0_tnlo}
\end{figure}

\begin{figure}[tbp]
\includegraphics[scale=0.33]{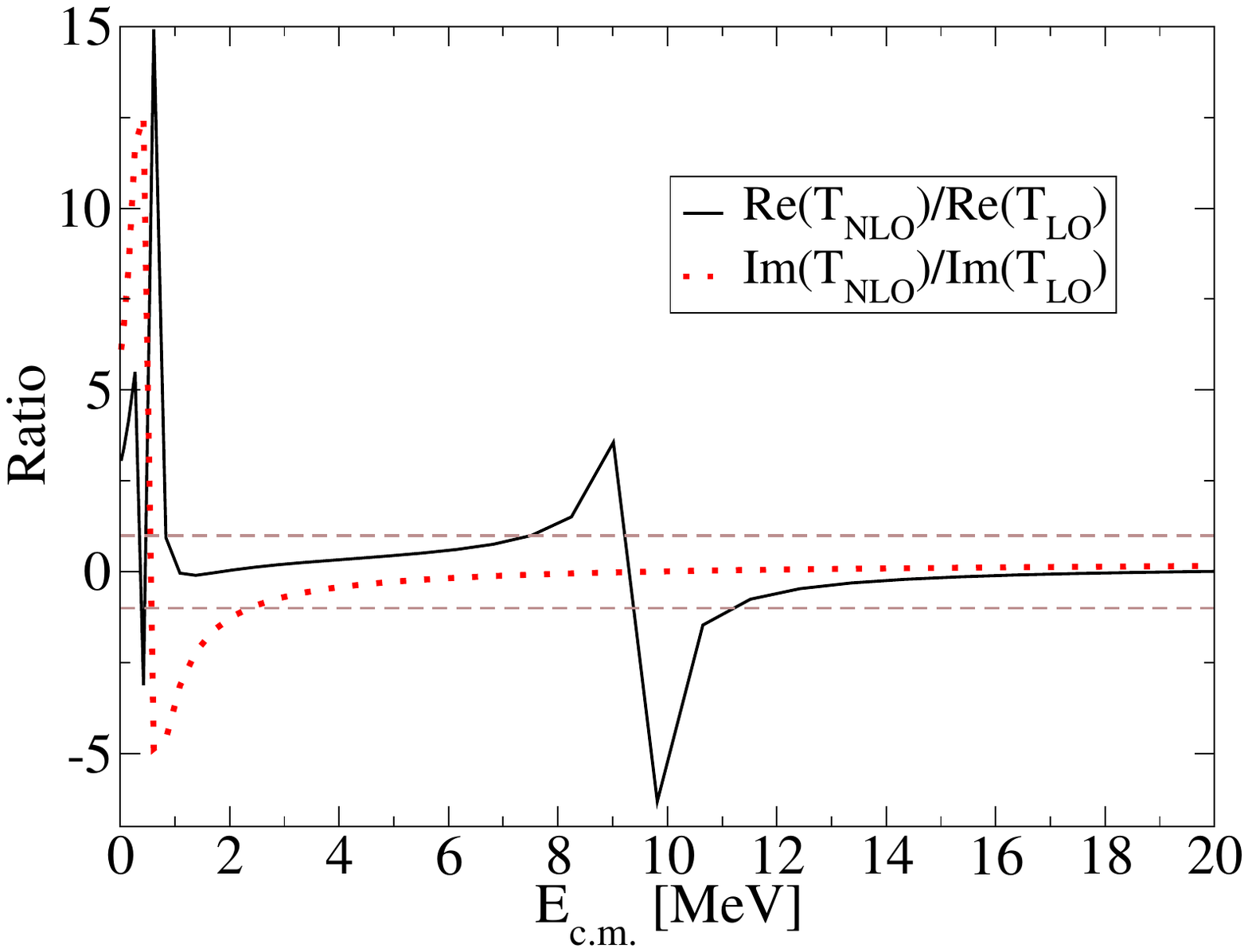}
\caption{The ratio of the real and imaginary part of the LO and NLO T-matrix as presented in Fig.~\ref{3p0_tnlo}. The two brown dashed-lines highlight the ratio$=1$ and $-1$.}
\label{3p0_r}
\end{figure}

Furthermore, one could include $c_{nlo}=-6.0\times10^{-10}$ MeV$^{-4}$---which is the value previously adopted in Table~\ref{t0}---and evaluate its perturbative correction. As shown in Fig.~\ref{3p0_s}, the results obtained by two different methods again agree with each other. However, the results appear to be unphysical for $E_{c.m.}=0.4-2$ MeV. Note that the LSE results are obtained by evaluating Eq.~(\ref{eqn:LSE23}) and Eq.~(\ref{nlophase}), while the HO-basis results are extracted by multiplying a small coefficient to $V_{NLO}$, diagonalizing the Hamiltonian and converting the eigenvalues and eigenfunctions via the J-matrix method, and finally extracting the perturbative phase shift $\delta_{NLO}$ by Hellmann–Feynman theorem~\cite{Yang:2016brl}. A check of the real and imaginary parts of $T_{LO}+T_{NLO}$ reveals that the unitarity of the S-matrix up to NLO is severely violated in the problematic energy domain, but the agreement between the two results suggests that the unphysical phenomenon is not due to numerical error or artifact of the HO-basis. 
In Fig.~\ref{3p0_tnlo} and Fig.~\ref{3p0_r}, the corresponding T-matrix at LO and NLO are plotted and compared. As one can see, the value of $T_{NLO}\gg T_{LO}$ for $E_{c.m.}=0.4-2$ MeV, and peaked around $E_{c.m.}\approx E_R\approx0.5$ MeV. Although in this region perturbation theory is not to be trusted and it could be meaningless to discuss the physical implications, it confirms the previous finding that shifting poles perturbatively around the threshold can be problematic.   

From the above check, it is safe to conclude that the main concern which could invalidate the analysis utilizing Eq.~(\ref{lo})-(\ref{10}) as presented in Sects.~\ref{3p0} and \ref{1s0}---i.e., the doubt of using HO-basis to represent the scattering states---can be alleviated by decreasing $\hbar\omega$ (to $\lesssim 1$ MeV) and increasing $n_{max}$ (to $\gtrsim100$). Since $\hbar\omega=1$ MeV and $n_{max}=150$ are adopted in Sects.~\ref{3p0} and \ref{1s0}, it is likely that the aforementioned problems concerning threshold crossing do not originate from the effects of the chosen HO-basis. 

\end{document}